\documentclass[1p]{elsarticle}
\usepackage{epsfig,amsmath,amssymb,amsfonts}
\usepackage{bm}
\usepackage[usenames]{color}

\newcommand\al{\alpha}
\newcommand\be{\beta}
\newcommand\de{\delta}

\newcommand\De{\Delta}
\newcommand\Ga{\Gamma}

\newcommand\la{\lambda}

\newcommand\half{{\frac{1}{2}}}

\newcommand\ce{{\cal E}}

\newcommand\ep{\epsilon}
\newcommand\vep{\varepsilon}
\newcommand\MD{\mathfrak{D}}
\newcommand\BMD{\bar{\mathfrak{D}}}
\newcommand\lam{\lambda}
\newcommand\tal{{\tilde{\alpha}}}
\newcommand\vak{\varkappa}

\newcommand\rmd{{\rm d}}

\begin{document}

\begin{frontmatter}

\title{Wave-packet continuum discretization for quantum scattering}

\author{O.A. Rubtsova }%
\ead{rubtsova-olga@yandex.ru}
\author{V.I. Kukulin}
\ead{kukulin@nucl-th.sinp.msu.ru}
\author{V.N. Pomerantsev}
\ead{pomeran@nucl-th.sinp.msu.ru}

\address{%
Skobeltsyn Institute of Nuclear Physics, Lomonosov Moscow State University,
Leninskie Gory 1(2), 119991  Moscow, Russia}

\date{\today}

 \begin{abstract}
A general approach to a solution of few- and many-body scattering
problems based on a continuum-discretization procedure is described
in detail. The complete discretization of  continuous spectrum is
realized using stationary wave packets which are the normalized
states constructed from exact non-normalized continuum states.
Projecting the wave functions and all scattering operators like
$t$-matrix, resolvent, etc. on such a wave-packet basis results in a
formulation of quantum scattering problem entirely in terms of
discrete elements and linear equations with regular matrices. It is
demonstrated that there is a close relation between the above
stationary wave packets and pseudostates which are employed often to
approximate the scattering states with a finite $L_2$ basis. Such a
fully discrete treatment of complicated few- and many-body
scattering problems leads to significant simplification of their
practical solution. Also we get  finite-dimensional approximations
for complicated operators like effective interactions between
composite particles constructed via the Feshbach-type projection
formalism. As illustrations to this general approach we consider
several important particular problems including multichannel
scattering and scattering in the three-nucleon system within the
Faddeev framework.
\end{abstract}

\begin{keyword}
quantum scattering theory \sep discretization of the continuum \sep
Faddeev equations \sep multichannel scattering
\end{keyword}
 \end{frontmatter}
\tableofcontents
\section{Introduction}
The formulation and practical solution for
 scattering problems in few- and many-body systems entails on
many difficulties caused by several reasons: (i) complicated
boundary conditions, especially above the breakup thresholds; (ii)
separation of the proper physical solution from unphysical (e.g.
forbidden by symmetry requirements etc.) ones; (iii)  high dimension
of resulting integral or integro-differential scattering equations;
\ (iv) presence of complicated moving singularities in the kernels
of few-body scattering equations.

All these difficulties have resulted in the current situation in the
field when we are able to treat rather accurately many-body bound
states in very complicated systems consisting of tens or hundred
particles while we are still unable to treat accurately even three-
or four-body systems with Coulomb interactions, the scattering of
particles with inner degrees of freedom, e.g. the collision of a few
molecules in quantum theory of chemical reactions, or a triple
collisions of three- or four nuclei etc. Moreover, even a solution
of conventional three-nucleon scattering problem with realistic
interactions requires an extensive usage of  supercomputers
\cite{gloeckle12}. It hampers strongly the further progress in this
important area. Thus,  one can formulate an important goal: not only
to simplify greatly  the solution of some few- and many-body
scattering problems, but also to develop a new method in scattering
theory which would be quite in a spirit of the bound-state problem
solving. So that the whole treatment of few- and many-body
scattering would be close to the routine bound-state calculations in
quantum chemistry or in nuclear physics.

On this way, many $L_2$-type methods have been suggested and widely
developed in the past and even now they are actively progressing.
 Among these the following methods can be mentioned: the R-matrix type approaches
\cite{rmatrix}, the equivalent quadrature technique
\cite{eq_quad,heller} and the Stieltjes--Tchebyshev moment-theory
approach \cite{langhof,reinhardt}, the Lorentz integral transform
\cite{LIT}, the harmonic oscillator representation \cite{horse} and
the J-matrix approach \cite{papp}, the continuum-discretized
coupled-channel method (CDCC) \cite{piyadasa,kami,rawitscher}, the
convergent close coupling approach \cite{bray}, different
realizations of the complex scaling method \cite{Lazauskas} and
others (see also the recent review \cite{Lazauskas_rep}). All these
approaches are utilized rather actively in atomic, chemical and
nuclear physics.

In this paper  we suggest another general approach for solving a
few-body scattering problem based on the usage of special normalized
states, the so-called stationary wave packets (WPs) or
eigendifferentials. The concept of eigendifferentials was introduced
 long ago by pioneers of quantum physics: H. Weyl \cite{weyl_orig},
E. Wigner \cite{wigner}, H. Bethe \cite{bethe} and others to treat
non-normalizable continuum states (which do not belong to a Hilbert
space)  in a framework of the standard theory of Hermitian operators
in a Hilbert space. The idea of stationary wave packets turned out
to be very fruitful at early stages of the quantum theory
development (see e.g. the detailed discussion in the classical
textbooks \cite{messia,greiner_QM}).

In our previous papers \cite{KC,K2,K3,K4,Moro,KPR1,KPR2,KPRF,KPR_br}
we proposed  to use such normalized WP states as a very convenient
discrete basis, onto which one can project out all the scattering
operators and wave functions. Because WPs are related explicitly to
exact scattering wave functions, many complicated operators such as
free or channel resolvents have  analytical finite-dimensional
approximations in the WP representation. This leads immediately to a
replacement of the scattering operators by finite matrices, so that
 a solution of a few-body scattering problem is reduced to
solving simple matrix equations with regular matrix elements instead
of multi-dimensional integral equations with singular kernels in a
conventional formulation. In addition, our approach results in a
very effective numerical scheme for a practical solution. Thus, in
the present paper we report the main results of the general WP
approach and describe how to reformulate complicated singular
few-body scattering equations into a form which allows to solve them
straightforwardly on a serial personal computer in a rather short
time.

The work has the following structure. Section 2 is dedicated to the
description of the eigendifferentials (stationary wave packets)
 and their interrelation to conventional non-normalized
scattering wave functions, on the one hand, and discrete
pseudostates resulting from a standard diagonalization procedure for
the Hamiltonian matrix, on the other. Section 3 includes a detailed
description of different properties and features of the stationary
wave packets. In Section 4 we present a discrete version of the
scattering theory based on the WP representation for the scattering
operators and wave functions.  Section 5 is devoted to the
multichannel scattering while  the next Section 6 contains a general
description of the three-body wave-packet formalism. In Section 7
 a solution of the composite particle scattering off a target-nucleus
is discussed. A formulation of the Faddeev equations in a discrete
form and their practical solution are given in  Section 8. We
summarize our main results in the last Section 9. For the reader's
convenience we added the Appendix which contains explicit formulas
for the  channel resolvent eigenvalues in the WP representation.

\section{Discrete representation for the scattering states}
Let us introduce some notations.   We start  here with two-body
scattering problem in a stationary formulation. Assume that the
potential $v$ is local and spherically symmetrical, so that the
angular momentum is conserved and the total Schroedinger equation is
reduced to radial equations for a fixed value of the angular
momentum $l$. Below we will deal mainly with the radial parts of the
wave functions which justify the partial Schroedinger equation:
\begin{equation}
\left(\frac{1}{2\mu}\left[-\frac{d^2}{dr^2}+\frac{l(l+1)}{r^2}\right]+v(r)\right)
\psi^l(r,E)=E\psi^l(r,E), \label{SE}
\end{equation}
where we use units with $\hbar=1$ and $\mu$ is the reduced mass. We
will omit the index $l$ wherever possible until Section 6 and also
other indices related to spin, isospin etc. until the cases they are
to be detailed.

The operator in the  left-hand side of eq.~(\ref{SE}) is the
 total Hamiltonian  $h$, which can be
written in the form:
\begin{equation}
h=h_0+v,
\label{h1}
\end{equation}
where $h_0$ is a free motion Hamiltonian. Further we will write all
the relationships in an operator form using the Dirac notations,
without specifying the representation. It is essential that
Hamiltonians $h_0$ and $h$ are Hermitian operators (in an
appropriate Hilbert space) and have a simple continuous spectrum
$[0,\infty)$.

The total Hamiltonian $h$ may also have a discrete spectrum
describing bound states of the system. The corresponding bound-state
wave functions of $h$ are denoted as
$\{|\psi^b_n\rangle\}_{n=1}^{N_b}$ while the continuum functions
 are denoted as $|\psi(E)\rangle$. They satisfy
 the Schroedinger equation (an operator form of the
eq.~(\ref{SE}))
\begin{equation}
h|\psi(E)\rangle=E|\psi(E)\rangle \label{SE2}
\end{equation}
  and  the usual orthogonality condition
 \begin{equation}
 \langle \psi(E)|\psi(E')\rangle=\delta(E-E').
 \label{norm_delta}
 \end{equation}
 In our study, we also need the scattering wave functions
$|\psi_{q}\rangle=\sqrt{\frac{q}{\mu}}|\psi(E)\rangle$ normalized to
the delta-function of momentum  $q=\sqrt{{2\mu E}}$:
\begin{equation}
\label{dek} \langle \psi_{q}|\psi_{q'}\rangle=\de(q-q').
\end{equation}

\subsection{Description of a continuous spectrum in terms of
 eigendifferentials}
A description of continuum meets some  difficulty in a stationary
formulation of a quantum mechanical problem. As is well known,
continuum wave functions for the total and free Hamiltonians, $h$
and $h_0$, do not have a finite normalization unlike to bound-state
wave functions and thus they do not belong to a Hilbert space.
Strictly speaking, such non-normalizable states are not
eigenfunctions of $h$ and $h_0$ and some basic properties of
Hamiltonians (the Hermiticity property) or even an orthogonality of
continuum wave functions could not be proven in a normal manner with
such non-normalizable set \cite{wigner,messia,greiner_QM}.

Nowadays this formal problem is avoided by using the Dirac
delta-function (\ref{norm_delta}) and by generalization of a Hilbert
space to a rigged Hilbert space.  However, in the past, physicists
often employed another formal trick (see e.g.
\cite{wigner,bethe,messia,greiner_QM}) based on the Weyl's
eigendifferential concept developed when studying the spectral
theory for singular differential operators~\cite{weyl_orig}.

The idea is  to introduce  an interval with a small width $\Delta E$
for each value $E$ from the continuum  and then to construct the
so-called eigendifferential, i.e. integral of the continuum wave
function $|\psi(E)\rangle$ over the interval:
\begin{equation}
\label{edf} |\psi(E,\Delta E)\rangle=\int_{E}^{E+\Delta
E}dE|\psi(E)\rangle.
\end{equation}
It is easy to see that such state is normalizable   (due to an
integration) and belongs to a Hilbert space. In this way, one can
generalize the conventional definition of a state normalization: the
state $|\psi(E)\rangle$ is treated as normalizable if its
eigendifferential has a finite norm \cite{bethe}.

To treat a whole continuum, it should be divided into
non-overlapping intervals (i.e. it should be discretized). The
system of eigendifferentials  forms the orthonormal set
\cite{greiner_QM}:
\begin{equation}
\langle \psi(E',\Delta E')|\psi(E,\Delta E)\rangle=
\left\{\begin{array}{ll} \Delta E,&\mbox{ for the same intervals,}\\
0,&\mbox{ for different intervals}\\
\end{array}
\right. .
\label{deltanorm}
\end{equation}
 It is important to stress that the set of such eigendifferentials
give a diagonal representation for the Hamiltonian:
\begin{equation}
\label{eig_delta}
 \frac{\langle \psi(E',\Delta E')|h|\psi(E,\Delta E)\rangle}{\Delta E}
=\left\{\begin{array}{ll} E +\half \Delta E,&\mbox{ for the same intervals,}\\
0,&\mbox{ for different intervals}\\
\end{array}
\right. .
\end{equation}
Besides this, they have a finite overlap with initial
(non-normalized) states
 \begin{equation}
 \label{pere}
 \langle \psi(E')| \psi(E,\Delta E)\rangle =
 \left\{
 \begin{array}{ll}
 1,&E'\in(E,E+\Delta E),\\
 0,&E'\notin
 (E,E+\Delta E)\\
 \end{array}
 \right. .
 \end{equation}
A complete system of eigenfunctions for the Hermitian Hamiltonian
  $h$ in the eigendifferential formalism is built from bound-state wave functions
 $|\psi^b_n\rangle$ and eigendifferentials
 \cite{wigner,greiner_QM}. In other words, for an
arbitrary wave function $|\Phi\rangle$ one can write down the
following expansion:
\begin{equation}
\label{spectr_sum} |\Phi\rangle=\sum_{n=1}^{N_b}C_n|\psi^b_n\rangle
+\sum C(E)|\psi(E,\Delta E)\rangle,
\end{equation}
where the sum over an infinite but accountable set of
eigendifferentials has rather a symbolic meaning. However further
one can pass to a limit $\Delta E\to 0$ which leads immediately to a
conventional expansion of the arbitrary function $|\Phi\rangle$
  over the bound-states and  functions of the continuum:
\begin{equation}
\label{spectr_sum1} |\Phi\rangle=\sum_{n=1}^{N_b}C_n|\psi^b_n\rangle
+\int_0^\infty dE \tilde{C}(E)|\psi(E)\rangle
\end{equation}
The discrete sum  in eq.~(\ref{spectr_sum}) goes to the integral
expansion in eq.~(\ref{spectr_sum1}) over continuum wave functions
$|\psi(E)\rangle$ in the sense of a Riemann--Stieltjes integral
\cite{greiner_QM}. Also it can be  shown \cite{greiner_QM} that the
finite normalization condition  for eigendifferentials
(\ref{deltanorm}) leads just to a delta-function normalization of
continuum states (\ref{norm_delta}).

Today such a way of definition of the Hamiltonian eigenfunctions for
a continuous spectrum practically is not used, as was noted above.
But in our approach we use such eigendifferentials as
basis states to solve a scattering problem in a discrete
representation. Below  we demonstrate  that this ``classical'' way
is turned out to be extremely fruitful for practical calculations in
few-body scattering problems.

There is also the another very popular concept for a treatment of a
discretized continuum: the pseudostates. It will be demonstrated
below   that these two ways of discretization:  via
eigendifferentials and via pseudostates  are closely interrelated.

\subsection{Pseudostates and  $L_2$ discretization of a continuum}
 As is known,  pseudostates  arise in
an approximate treatment of a continuous spectrum of a quantum
system based on a  projection of scattering wave functions into an
orthogonal $L_2$  basis set $\{|\phi_n\rangle\}_{n=1}^N$ of a finite
dimension $N$.

If one substitutes the expansion of a Hamiltonian eigenfunction in
the basis functions  $|\psi\rangle=\sum_{n=1}^NC_n|\phi_n\rangle$
into the Schroedinger equation (\ref{SE2}) one gets the following
system of algebraic linear equations for the expansion coefficients
$\{C_n\}_{n=1}^N$
\begin{equation}
\sum_{k=1}^N [h_{nk}-E\de_{nk}] C_k=0,\quad n=1,\ldots,N,
\end{equation}
which corresponds to the eigenstate problem for the Hamiltonian
matrix with the elements
$h_{nk}\equiv\langle\phi_n|h|\phi_k\rangle$. After a diagonalization
of the Hamiltonian matrix one gets a  set of
eigenvalues\footnote{Below we denote eigenvalues by an upper
asterisk index to distinguish them from discretization interval
end-points.} $\{\epsilon_n^*\}_{n=1}^N$ and eigenfunctions
$|\psi_n\rangle=\sum_{k=1}^NC_k^n|\phi_k\rangle$ $(n=1,\ldots,N)$.
This discrete set can be divided on two groups: the eigenfunctions
with lower eigenvalues describe  the system bound states if they
exist (in the case of a short-range potential their number is
finite) while the remained  eigenstates correspond to the continuum
wave functions at discrete energies $\epsilon_n^*$. Unlike to the
exact continuum functions these discrete eigenstates have a finite
norm and their eigenvalues define the discretized continuum. That is
why they are usually referred to as {\em  pseudo}states of the
Hamiltonian $h$.

For the above pseudostates, the conditions similar to those given in
eqs.~(\ref{deltanorm}-\ref{eig_delta}) for eigendifferentials are
 valid:
\begin{equation}
\label{eig_pseu}
\begin{array}{c} \langle \psi_n|\psi_k \rangle =\de_{nk},\\
\langle \psi_n|h|\psi_k\rangle=\epsilon_n^*\de_{nk}\\
\end{array}
\end{equation}
An employment of  pseudostates instead of  exact scattering functions in
practical scattering calculations meets the well known difficulty related to the
difference in their normalization. The functions of pseudostates  behave in some inner region like
the scattering wave functions but, in
sharp contrast with the latter, they vanish in the asymptotic region. Nevertheless, the
functions of pseudostates together with bound-state wave functions form a
complete set of basis functions, and this set can be used to approximate the
spectral expansions of the scattering operators or scattering wave functions.

\subsection{An equivalent quadrature concept}
Very often it is necessary to find not  operator itself, but  only its matrix elements
with normalized functions $|\Phi\rangle$. The example for
such an employment is the calculation of transition probabilities
between the bound states of subsystem via the intermediate continuum
states, in particular,  the calculation  of response functions in the Lorentz
integral transform (LIT) method \cite{LIT} or the calculation of
$t$-matrix elements on the basis of spectral expansion of the total
resolvent of the system~\cite{reinhardt}.

Let us consider a matrix element of some operator $B=f(h)$ which is
a function
 of the Hamiltonian $h$ using the spectral expansion
of the operator $B$  in a complete set of the Hamiltonian
eigenfunctions:
\begin{equation}
\label{phi_a} \langle \Phi|B|\Phi\rangle=\sum_{n=1}^{N_b}
f(E_n)|\langle \Phi|\psi^b_n\rangle|^2+\int_0^\infty dE f(E)|\langle
\Phi |\psi(E)\rangle|^2.
\end{equation}
As is evident, the spectral expansion (\ref{phi_a}) includes not
wave functions themselves but their positive  quadratic forms
 $|\langle \Phi|\psi (E)\rangle|^2$.
So that, for such quantities there are no problems related to the
normalization of scattering wave functions and they are finite
functions of energy. Further, the integral over continuous
spectrum can be approximated by a discrete sum over respective
pseudostates:
 \begin{equation}
\langle \Phi|B|\Phi\rangle\approx\sum_{n=1}^{N_b} f(E_n^*)|\langle
\Phi|\psi^b_n\rangle|^2+\sum_{n=N_b+1}^N f(E_n^*)|\langle \Phi
|\psi_n\rangle|^2.
\end{equation}
Thus, such a ``spectral expansion'' over pseudostates can be treated
as some effective quadrature for the integral over continuum in
eq.~(\ref{phi_a}). This approach has been suggested earlier for
solution of different particular problems in atomic physics
\cite{eq_quad,heller,langhof,reinhardt} related to finding matrix
elements of Green functions. E.g.,  a conception of an equivalent
quadrature (EQ) was introduced within the framework of the
moment-theory approach for calculation of non-negative spectral
densities \cite{eq_quad,langhof}.
  The mesh points  of the EQ coincide with
pseudostate energies and corresponding weights are determined
formally from relations between discrete and continuous quantities:
 \begin{equation}
 \label{eq_def}
 |\langle \Phi|\psi_n\rangle|^2=\omega_n |\langle
 \Phi|\psi(\epsilon_n^*)\rangle|^2, \quad n=N_b+1,\ldots,N
 \end{equation}
 which are valid at discrete pseudostate energies
$\epsilon_n^*$.

One important result of the equivalent quadrature technique is the
fact the quadrature weights do not depend upon the function
$|\Phi\rangle$ and are determined actually by pseudostate wave
functions and by the spectral density of the particular system.
 So that, the values  $\omega_n$ are just
the coefficients for transitions from normalizable states
$|\psi_n\rangle$ to non-normalizable functions
$|\psi(\epsilon_n^*)\rangle$.


In general, finding the weights of the equivalent quadrature is highly
non-trivial problem because the exact scattering wave functions
  $|\psi(E)\rangle$ are still unknown. However, for many problems when one
  needs only complete spectral sum, knowledge of these weights is not required.
For example, the pseudostate functions are used for inclusion of
  intermediate continuum states for excitations of projectile or target
   in description of
  composite particle scattering in atomic
 \cite{bray} and nuclear
 \cite{piyadasa} physics. In these cases, as a rule, the pseudostates are employed in
expansion of a total wave function of the system to perform a
coupled channel reduction of the initial many-body problem.

However, it is still possible  to find scattering observables just from
individual pseudostate wave functions if the EQ is known. In fact,
if to write down an expression for the elastic scattering amplitude:
\begin{equation}
A(E)=\langle \psi_0(E)|v|\psi(E)\rangle,
\end{equation}
it is easy to see that the function  $v|\psi_0(E)\rangle$, where
$|\psi_0(E)\rangle$ is a free motion function, is square-integrable
in case of a short-range potential $v$. So that, one can employ the
properly normalized pseudostate instead of the exact scattering wave
function:
\begin{equation}
\bar{A_n}\approx\langle {\psi}_{0}(\epsilon_n^*)|v|\psi_n\rangle.
\end{equation}
To find direct relation between ``continuous'' $A(E)$ and
``discrete'' $\bar{A_n}$ amplitudes, one can use the EQ definition
(\ref{eq_def}) which results in:
\begin{equation}
\label{aom} A(E_n)\approx \frac{\bar{A_n}}{\sqrt{\omega_n}}.
\end{equation}
Actually, the problem of evaluation of $\omega_n$  is related
to finding a spectral density in the continuous spectrum of
$h$. One of the methods is to evaluate moments of the energy
distribution using spectral expansion in
pseudostates. Then, the equivalent quadrature weights can be found
from a solution of complicated set of nonlinear algebraic equations
\cite{reinhardt}. However, to find the EQ, the eigendifferential
formalism can be used.
\subsection{Pseudostates as approximations for eigendifferentials}

In our previous works \cite{K2,K3,KPR1} the following way for
finding the coefficients of transformation from normalizable
functions to non-normalizable ones has been suggested. This method
is based on the observation that pseudostate wave functions are very
similar to eigendifferentials. In particular, both types of
functions behave in the inner region  like an exact scattering wave
function, whereas in the asymptotics they vanish differently in
accordance with their own properties. Moreover, functions from both
sets are normalized and the Hamiltonian matrices in both sets are
diagonal.

It means one can treat  pseudostates as approximations just for
eigendifferentials rather than for exact non-normalizable scattering
wave functions\footnote{It is worth noting that the interrelation of
pseudostates and eigendifferentials has been also proved in
refs.~\cite{piyadasa,kami} on the basis of the CDCC method. In this
approach
 these two types of states are considered  to treat a composite
particle scattering  off heavy target. However,  direct relations
 between them have been  established for the equidistant discretization
distribution only.}. Of course, this approximation is valid only
when eigenenergies of both states (\ref{eig_delta}) and
(\ref{eig_pseu}) are coincided.

\begin{figure}[h!]
\centering\epsfig{file=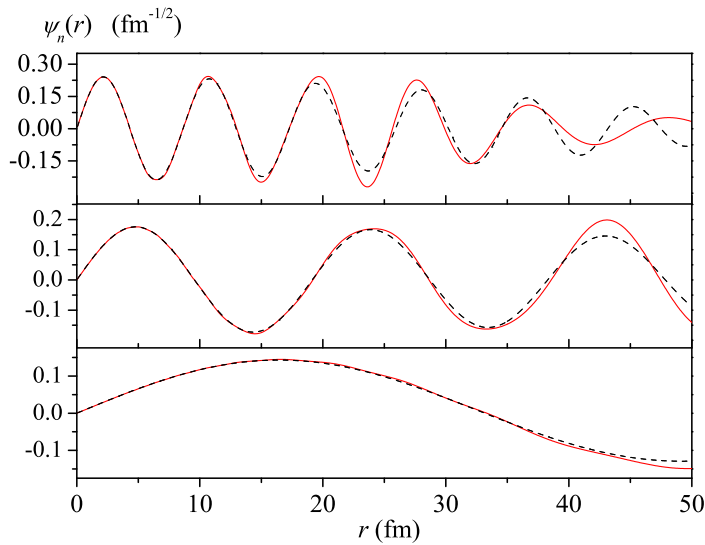,width=0.8\columnwidth}
\caption{\label{eigen} Comparison of  pseudostates (solid curves) of
the free motion $NN$ Hamiltonian obtained in a basis of Gaussians
with exact eigendifferentials (dashed curves) constructed for the
same distribution of eigenenergies of the free Hamiltonian
discretized continuum.}
\end{figure}

In  Fig.\ref{eigen} we compare pseudostates of the free Hamiltonian
found in the Gaussian basis (see details in ref.~\cite{KPR1}) with
 exact eigendifferentials (normalized to unity) constructed for
the same distribution of eigenenergies $\ep_n^*$. It is clear that
both type functions are nearly coincide  in some inner domain.

Further,  it is easy to see that the coefficients of transformation
to non-normalizable states (see the definition~(\ref{edf})) are
given simply by the widths $\Delta E$ in the
discretization procedure for eigendifferentials:
\begin{equation}
\label{ade} A(E)\approx\frac{\langle \psi_0(E)|v|\psi(E,\Delta
E)}{{\Delta E}}.
\end{equation}
Using the direct correspondence between eigendifferentials (with the
normalization factor) and pseudostates
 \begin{equation}
\frac{\langle \psi_0(\ep_n^*)|v|\psi(\ep_n^*,\Delta
E_n)\rangle}{\sqrt{\Delta E_n}}\approx {\langle
\psi_0(\ep_n^*)|v|\psi_n\rangle},
\end{equation}
and comparing the eqs.~(\ref{aom}) and (\ref{ade}), one gets an
approximate relation:
\begin{equation}
\omega_n\approx\Delta E_n.
\end{equation}
So that, we find an amazing result: the coefficients of
transformation from pseudostates to exact continuum wave functions
are determined, first of all, by the distribution of discrete
pseudostates energies $\epsilon_n^*$ rather than by the form of
basis functions $|\phi_n\rangle$ themselves.  Actually the form of
basis functions determines implicitly the pseudostate energy
distribution: it  will be different  for various  basis sets.

It would be very appropriate to give here some clear example. Let us
consider a free particle motion in a spherical box with radius $R$.
The solution of a radial Schroedinger equation for $S$-wave in this
case is an infinite set of wave  functions (normalized to unity)
\begin{equation}
\psi_{0n}(r)=\sqrt\frac2R\sin q_n^*r,\quad q_n^*=\frac{\pi
n}{R},n=0,\ldots,
\end{equation}
instead of exact free motion states:
\begin{equation}
\psi_{0q}(r)=\sqrt{\frac2{\pi}}\sin qr.
\end{equation}
For this case the weights of the EQ do not depend on the index $n$
and can be found explicitly:
\begin{equation}
\omega_n=\frac{\pi}{R}.
\end{equation}
It is clear that this quantities coincide {\em exactly} to the
discrete momentum differences $d_n=q_{n+1}^*-q_n^*$ (momentum analog
of the energy widths in (\ref{edf})). In other words, one can
determine the weights of the EQ directly via the discretization
width
\begin{equation}
\omega_n=d_n
\end{equation}
rather that via the size of the box $R$ (although they are
interrelated).

When we are dealing with an  $L_2$-discretization using
an arbitrary basis, this can be treated as a solution of the
scattering problem in a box with blurred boundaries but  these
boundaries are difficult to determine explicitly. However they do
impact on a discrete energy distribution which is used for finding
approximate weights or widths of discretization intervals. So that,
eventually in order to derive the values of scattering observables
from the discrete energy distribution one should determine the spectral
partition parameters using eigenvalues
 $\epsilon_n^*$ of pseudostates obtained from the Hamiltonian matrix
 diagonalization. In general case, the  energy interval
  widths $\Delta E_n$ would be different for different
 $n$.

Thus, our main idea here is that pseudostates are approximations for
exact  eigendifferentials. Surely, the above argumentation is not
strict but plausible. It is also necessary to have similarity
between pseudostates and eigendifferentials at least within the
interaction range. It means that the size of the pseudostate basis
should be rather large and the interval widths to be rather small.
Unfortunately, it is difficult to give more strict conditions for
such a replacement here. Some estimates can be found in
ref.~\cite{KPRF}.

Below we formulate a fully discrete version of a stationary
scattering theory  using the eigendifferential (wave-packet)
formalism, while  in practical realizations we  replace often exact
eigendifferentials with  pseudostates found from the Hamiltonian
matrix diagonalization on an appropriate basis. Both these points
make it possible  to  simplify greatly the practical solving of
few-body scattering problems.

\section{Stationary wave packets and their properties}
In the approach which will be used everywhere in this work, the
eigendifferentials will be employed as primary basis functions
upon which all wave functions and operators will be expanded. In
other words, we will treat scattering problems in a discrete
representation.

We  rename eigendifferentials as  stationary wave packets because
the name seems to us more physical and clear for understanding.
The approach developed
--- we refer to it as the wave-packet continuum discretization method
(WPCD) ---  takes  the advantages of the above $L_2$ techniques and
simultaneously gives an accurate  solution  of  few-body equations.

\subsection{Basic definition of stationary wave packets}
 It is important to stress that the stationary wave packets are constructed from
exact wave functions of continuous spectrum as their
eigendifferentials (normalized to unity). But WPs are considered
here as ordinary functions and thus the details of a partition
procedure for a continuous spectrum have to be defined.

 Let us restrict the continuous spectrum of the free Hamiltonian $h_0$ $[0,\infty)$
 with some maximal
 value of energy $E_{\rm max}$ and then divide the whole interval
$[0,E_{\rm max}]$ into finite number of non-overlapping intervals
 $[\ce_{i-1},\ce_i]_{i=1}^{N}$ (where $\ce_0=0$ and
$\ce_N=E_{\rm max}$) which we recall, according to
ref.~\cite{piyadasa}, as discretization bins. We assume here that
the value  $E_{\rm max}$ is finite although sufficiently large in
order to provide convergence to the exact solution of the problem to
be treated. Every such an energy bin corresponds to the  interval
$[q_{i-1},q_i]$ on the momentum axis $q$ where $q=\sqrt{{2\mu E}}$
and $\mu$ is the reduced mass. For the sake of convenience we will
denote both the momentum and the energy bins by a symbol   $\MD_i$
($i$ is an interval number). Also we denote widths of momentum and
energy intervals as follows:
 \begin{equation}
 d_i=q_i-q_{i-1},\quad D_i=\ce_i-\ce_{i-1}.
\end{equation}
Further we consider the complete set of continuous spectrum states
 $|\psi_{0q}\rangle$
for the free  Hamiltonian  $h_0$
 (for each partial wave $l$) normalized as in eq.~(\ref{dek}).
Now we define the free stationary wave packets  as integrals of free
motion wave functions over the discretization bins  as in
eq.~(\ref{edf}):
\begin{equation}
\label{ip} |x_i\rangle=\frac{1}{\sqrt{B_i}}\int_{\MD_i}dq
f(q)|\psi_{0q}\rangle,\quad i=1,\ldots,N,
\end{equation}
where $f(q)$ is some weight function and  $B_i$ is a normalization
coefficient related to each other by the formula:
\begin{equation}
\label{norm} B_i=\int_{\MD_i}dq|f(q)|^2.
\end{equation}
By modifying the weight function $f(q)$ one can get different types
of WPs. In particular, the case
 $f(q)=\sqrt{q/\mu}$ corresponds to the energy wave-packets which coincide
with eigendifferentials  (\ref{edf}) up to  normalization factors.
The case $f(q)=1$ leads to momentum WPs etc. However it is important
to note that when the  partition used is sufficiently dense (i.e.
for small bin widths) the particular form of the weight function
does not matter. Moreover in practical calculations one can replace
energy WPs with momentum ones (for the same partition) and  errors
related to this replacement are  very small.

According to the definition (\ref{ip}) it is easy to show that the
states $|x_i\rangle_{i=1}^N$ form an orthonormalized set   \cite{K2}:
\begin{equation}
\langle x_i|x_j\rangle=\de_{ij}, \quad i,j=1,\ldots, N.
\end{equation}
Then it is straightforward to get a formula for the overlap integral
between WP states and initial plane waves:
\begin{equation}
\label{projk} \langle x_i|\psi_{0q}\rangle=\left\{
\begin{array}{rr}
\displaystyle \frac{f(q)}{\sqrt{B_k}},&\quad q\in \MD_i,\\
0, &\quad  q>q_{N}.\\
 \end{array}
\right.
\end{equation}
The free  Hamiltonian matrix  (and also the matrix of any operator
depending upon $h_0$ or just commuting with it) is {\em diagonal} in
such a basis while the respective eigenvalues depend on the weight
function:
\begin{equation}
\label{evh0} \langle
x_i|h_0|x_j\rangle=\frac{\delta_{ij}}{B_i}\int_{\MD_i}|f(q)|^2\frac{q^2}{2\mu}dq.
\end{equation}
In the case of energy WPs, the eigenvalues of $h_0$ are just the bin
midpoints:
\begin{equation}
\langle x_i|h_0|x_i\rangle\equiv\ce_i^*=\half(\ce_{i-1}+\ce_i),
\end{equation}
and in the case of momentum WPs, the eigenvalues are:
\begin{equation}
\ce_i^*=\frac{1}{6\mu}(q^2_{i-1}+q_{i-1}q_i+q_i^2).
\end{equation}

Such WP states will be used to carry out all our calculations. From
practical point of view the wave-packet set can be used in any
scattering calculations on the same footing as usual discrete  $L_2$
bases like  the harmonic oscillator basis or a basis of Gaussians.
However, in contrast to those, such WP basis states have a direct
relation (\ref{projk}) to exact continuum wave functions.
\subsection{Behavior of wave-packet functions in configuration and momentum spaces}
The WP  functions have an interesting behavior both in coordinate
and momentum spaces. Consider for simplicity the case of
 $f(q)=1$. It is easy to show \cite{K2,KPR1} that in coordinate space
WP functions behave as follows (here we use the reduced radial part
of a  wave function for $s$-wave):
\begin{equation}
 \label{exact}
 x_i(r)= \sqrt{\frac{2d_i}{\pi}}\sin(q_i^*r)\frac{\sin({d_ir}/{2})}{{d_ir}/{2}},
\end{equation}
where $q_i^*=\half(q_{i-1}+q_i)$ is a midpoint of the interval
$\MD_i$. So that, free WPs are nearly coincide with non-normalized
continuum wave functions in the area
    $ r\ll r_i=\frac2{d_i}$.
As is seen from eq.~(\ref{exact}), the ``packetting'' procedure
defined in eq.~(\ref{ip}) results in an additional decreasing factor
 which just provides the wave function with  finite
normalization.

\begin{figure}
\centering \epsfig{file=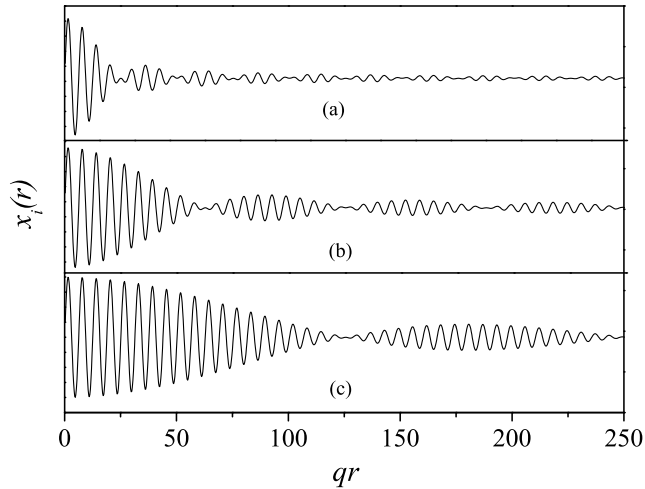,width=0.8\columnwidth}
\caption{\label{fig2_1} Coordinate behavior of $s$-wave WP states
$x_i(r)$ for different values of bin widths in ratio to momentum
value $q$: $d_i/q=0.25$ (a), $d_i/q=0.1$ (b) and $d_i/q=0.05$ (c).}
\end{figure}

In Fig.~{\ref{fig2_1}} the coordinate behavior of three WP states
(\ref{exact}) are compared for different ratios of bin width   $d_i$
and its midpoint value  $q_i^*$. It follows from the Figure that the
radial behavior of the WPs resemble the characteristic picture of
beats (i.e. dual-frequency oscillations) and the smaller ratio
$d_i/q_i^*$ corresponds the slower decay of the WP function.
Therefore such a basis functions can be employed to approximate the
non-normalized continuum wave functions on longer distances.
This property is crucially important for few-body scattering
calculations because  the Faddeev components of the total wave function
have a specific long-range asymptotic behavior which
must be reproduced correctly.

 In the momentum representation, the free WP
functions are simple step-like functions: they do not vanish only
for the momentum interval coinciding with their bin, i.e. only for
the on-shell region. In this area they are fully determined by the
weight function $f(q)$:
\begin{equation}
\label{step}
x_i(q)=\frac{f(q)(\theta(q-q_{i-1})-\theta(q-q_i))}{\sqrt{B_i}},
\end{equation}
where the $\theta$ is the Heaviside step-like function\footnote{Hear
the $\theta$-functions is defined as follows:
$$\theta(q)=\left\{
\begin{array}{lr}
 1,&q\geq0\\
  0,& q<0\\
\end{array}
\right..$$ }. When $f(q)=1$ all the wave functions in the momentum
WP representation take a histogram form. Being generalized onto a
few-body case, the free few-body WP basis functions are built as a
direct products of step-like functions along every independent
momentum variable. In this way, the whole momentum space is replaced
by a finite momentum lattice. In this sense we will refer to this
basis as {\em a lattice basis}.

\subsection{Wave packets as a basis}
For further investigations, the important questions are: does the
WP set form a basis and  how one can prove  a convergence of the
results with increasing a basis dimension. It should be noted here that
for the set of pseudostates the problem of  completeness does not
exist because
 the pseudostates jointly with the bound states form the same linear
span as the initial $L_2$ basis set. So that, the set of
pseudostates and bound states form a basis which becomes to be
complete
 in the limit $N\to\infty$.

It is clear that, unlike to the eigendifferential case when $\Delta
E\to 0$, the WP set does not form a complete basis in a full Hilbert
space if bin widths are kept to be finite. But one can introduce the
WP subspace ${\cal H}_P$ in which the orthonormal set
$\{|x_i\rangle\}_{i=1}^N$, of course, forms a basis. The projector
 onto the WP subspace is defined as usual:
\begin{equation}
\label{projector} \mathfrak{p}=\sum_{i=1}^N|x_i\rangle\langle x_i|.
\end{equation}
Thus below  we will refer to the WP set  as a WP basis keeping in
mind   it is the basis in ${\cal H}_P$.

One has to make some estimates concerning the convergence of  the
results obtained in ${\cal H}_P$ to exact scattering-problem
solutions with increasing the basis dimension.

 The exact unit operator $I$, which can be written in the following form:
\begin{equation} \label{total_i} I\equiv \sum_{i=1}^{N} I_i+I_{\rm
r}=\sum_{i=1}^{N}\int_{\MD_i} |\psi_{0q}\rangle \langle\psi_{0q}|
dq + \int_{q_{N}}^\infty |\psi_{0q}\rangle \langle\psi_{0q}| dq,
\end{equation} is replaced in the WP approach  with the
wave-packet projection operator $\mathfrak{p}$. This replacement
implies two following approximations:
\begin{itemize}
 \item[(i)] the infinite continuous spectrum is truncated with
the maximal momentum value $q_{N}$ and the residual integral
$I_{\rm r}$ is neglected;
 \item[(ii)] exact partial spectral
projectors $I_i$ are replaced with the WP partial projectors
$|x_i\rangle \langle x_i|$. \end{itemize}
Surely, these points are
not valid in a full Hilbert space. But keeping in mind numerical
applications of the method one can compare the mean values of
operators $I$ and $\mathfrak{p}$ in some $L_2$ normalized state
$|\Phi\rangle$ having an effective range $r_0$. To satisfy to the
following conditions (corresponding to the above points (i) and
(ii))
\begin{equation} \langle\Phi|\mathfrak{p}|\Phi\rangle
\approx\langle\Phi|\sum_{i=1}^{N} I_i|\Phi\rangle, \quad
\langle\Phi|I_r|\Phi\rangle \ll \langle\Phi|\sum_{i=1}^{N} I_i|\Phi\rangle
\end{equation} one has to have sufficiently small widths $d_i$ and sufficiently
high maximal momentum  value $q_{N}$ \cite{KPR1}: \begin{equation}
\label{complete} d_i\ll \frac1{r_0}\ll q_{N},\quad i=1,\ldots,N.
\end{equation} With these restrictions one can choose for the
practical realization the proper conditions for the momentum bin
partition $\{\MD_i\}_{i=1}^{N}$. Then one can check the
convergence of the results with increasing a wave-packet basis
dimension, when the bin widths become smaller and smaller, and the
maximum value $q_{N}$ becomes higher  and higher. This procedure
 simulates  changes in eigenvalue $\{\ep_n^*\}$  distributions for
  pseudostates  with increasing a basis dimension $N$.

One can employ, for example, distributions (grids) transforming a
finite interval to whole numerical axis, e.g. the Tchebyshev grid
\cite{KPR2}):
\begin{equation}
q_i=q_{\rm m}\tan\left[\frac{2i-1}{4N}\pi\right],\quad i=1,\ldots,N,
\end{equation}
where $q_{\rm m}$ is a scale parameter.

Such grid points $q_i$ are convenient because one can study the
convergence of the results with increasing the basis dimension as in
the usual $L_2$-basis case. One can expect that  the result obtained
in the WP representation approaches  the exact solution with
increasing the basis dimension $N$.

\subsection{Construction of WP states for the total Hamiltonian}
Similarly to the free Hamiltonian  ($h_0$) case, one can introduce a
WP set for the total Hamiltonian $h=h_0+v$.  We define a partition
of the continuum for the total Hamiltonian  into non-overlapping
bins $\{\Delta_k=[\epsilon_{k-1},\epsilon_k]\}_{k=1}^M$ with
$\ep_0=0$ and $\ep_M=E_{\rm max}$ (or $\Delta_k=[q_{k-1},q_k]$).
Let's note that this partition might be different from the free WP
partition $\{\MD_i\}$. The {\em scattering wave-packets} are defined
as integrals of exact scattering wave functions $|\psi_q\rangle$
over the intervals chosen:
\begin{equation}
\label{zi} |z_k\rangle=\frac1{\sqrt{C_k}}\int_{\De_k}{\rm
 d}q\  w(q)|\psi_q\rangle,
\end{equation}
where $C_k$ and $w(q)$ are the normalization coefficients and weight
functions respectively. They have the same properties as free WPs.
However, the total Hamiltonian may have a discrete spectrum. In this
case  one has to add  bound-state wave functions of $h$ to the
scattering WP set to obtain  complete WP set corresponding to the
total Hamiltonian $h$.

In practical solution we have the basic problem how to construct
these scattering WPs without solving the initial scattering problem.
Here one can exploit an idea of similarity of eigendifferentials
(WPs) and pseudostates. So that, one can use  free WPs as a basis to
construct scattering WPs for the total Hamiltonian $h$ as its
pseudostates.
 For this purpose we  apply  a diagonalization procedure for the total
Hamiltonian matrix in a free WP basis. As a result we get a
discrete sets of eigenvalues $\ep_k^*$ and respective eigenvectors
$|\tilde{z}_k\rangle$.

\begin{figure}[h]
\centering \epsfig{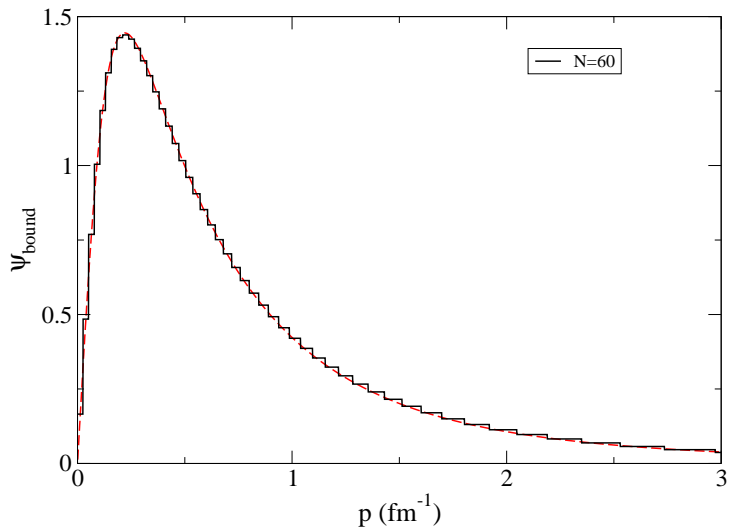}
 \caption{\label{fig0}  Comparison of the exact deuteron wave
 function in the momentum representation for the Yamaguchi potential
 (dashed curve) with its approximation in the lattice basis (solid line).}
\end{figure}

 Since the free WP basis functions (in the momentum
space) are step-like  functions, the  momentum dependence of all
functions expressed via such a basis have a histogram-like form. An
example of the momentum dependence for the bound state (deuteron)
function in such step-like basis in comparison with the exact
function for the Yamaguchi $s$-wave triplet $NN$ potential  is
displayed in Fig.~{\ref{fig0}}.

All the eigenstates corresponding to the area of continuous spectrum
obtained in such a WP-representation can be treated as
approximations for scattering wave packets $|z_k\rangle$ defined
above, provided the
  eigenvalues of Hamiltonian matrix coincide with eigenvalues of
the exact wave-packets. This condition can be easily satisfied
because one can construct a bin partition for exact scattering WPs
``by hands".

\begin{figure}[h]
\centering \epsfig{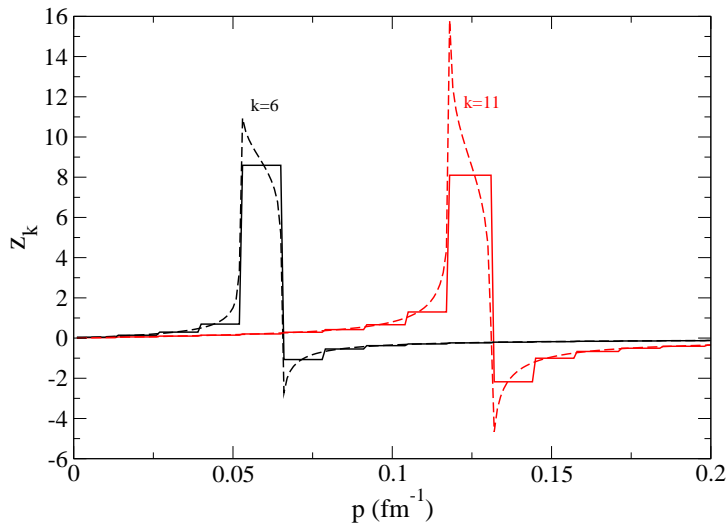}
\caption{\label{wpac} (Color online) The functions of pseudostates
($k=6,11$) obtained in the lattice basis (solid lines) in comparison
with exact scattering wave-packets (dashed lines) for the  $NN$
spin-triplet Yamaguchi potential. } \end{figure}

As a result, we find very convenient discrete representation for
the scattering WPs using the free wave packets: \begin{equation}
\label{rot1}
|z_k\rangle\approx|\tilde{z}_k\rangle=\sum_{i=1}^NO_{ki}|x_i\rangle.
\end{equation} The validity of this approximation is illustrated  by the same
particular example of the Yamaguchi potential for which both exact
solutions and also exact scattering WPs are constructed in an
explicit form and thus one can compare them with approximations
given by eq.~(\ref{rot1}). In Fig.~{\ref{wpac}} we display the
functions of two pseudostates in momentum representation (with indices $k=6$ and 11) obtained
in the free WP basis in comparison with the corresponding exact
scattering wave packets. It is interesting that the exact scattering
WPs (\ref{zi}) (dashed curves in the Figure) are square-integrable
functions, despite the fact that  they have logarithmic
singularities  on the boundaries of the ``on-shell'' interval, i.e.
the interval  to which the WP energy $\ep_k^*$ belongs.
  It is clear from
the above comparison that the pseudostates composed from step-like
WPs reproduce quite reasonably the structure of exact scattering
wave packets ``on average''.

Such a  discrete approximation for the scattering WPs is extremely
important for few-body scattering studies where one is able to build
a few- and many-body WP basis not only for a free motion Hamiltonian
but also for a few-body channel Hamiltonian. Moreover, in this
approach  one can find a very convenient analytical form for the
resolvent and off-shell $t$-matrix operators for a two-body
subsystem by a straightforward Hamiltonian diagonalization
procedure.

Furthermore we will demonstrate henceforth how to generalize this
convenient way to the general multichannel scattering. However, in
general case of a multi-particle scattering one has to take into
account a specific behavior of scattering wave functions in
different asymptotic channels. In such cases we will formulate
discrete versions for the Faddeev and many-body Lippmann--Schwinger
integral equations, see Sections 6-8.

\subsection{Coulomb wave packets}
One of the big advantages of an employment of $L_2$ normalized WP
states for a continuum treatment is a possibility to consider
long-range potentials like the Coulomb one quite similarly to
short-range ones. It is because the essential singularities at small
momentum $q$ peculiar to wave functions and transition operators for
the Coulomb potential are averaged out and smoothed in the WP
representation.

As an illustration let us consider two-body scattering with
repulsive Coulomb Hamiltonian\footnote{The Coulomb attraction can be
also treated in the WP approach, however it needs in a separate
study.}
 \begin{equation}
 \label{ham_c}
 h_C=h_0+\frac{z_1z_2e^2}{r},
 \end{equation}
where $z_1$ and $z_2$ are the particle charges, $r$ is the distance
between them. To treat the case of charged particles one can employ
the Coulomb wave packets  $|x_i^C\rangle$ as the basis functions.
Such Coulomb WPs may be built from the regular Coulomb wave
functions $F_l(q,r)$ by the integration over discretization bins
\cite{KC} quite similarly to the general case:
\begin{equation}
\label{x_coul} |x_i^C\rangle=\frac1{\sqrt{B_i}}\int_{\MD_i}{\rm
 d}q f(q)|F_l(q)\rangle.
\end{equation}
It can be shown straightforwardly that WP states (\ref{x_coul}) are
normalized  and thus one can construct them practically using
pseudostates of the Hamiltonian (\ref{ham_c}) on some $L_2$ basis.
Thus, in such a discrete representation one gets an interesting
picture when one can expand the Coulomb WPs over finite set of
free-motion WPs.

\begin{figure}[h]
\centering \epsfig{file=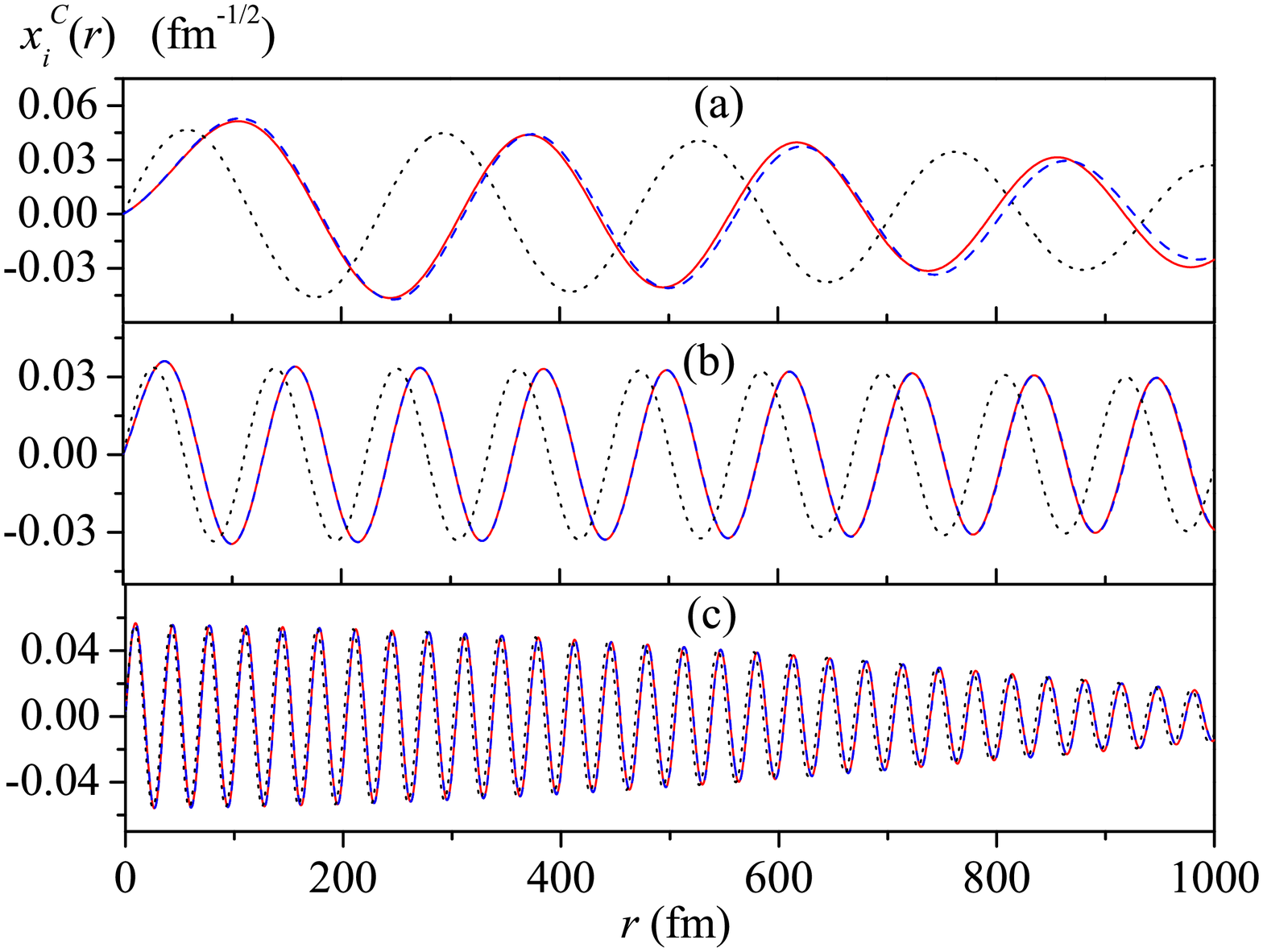,width=0.8\columnwidth}
\caption{\label{coul_fun} The exact Coulomb WPs (dashed curves), the
pseudostates found in the free WP basis (solid curves) and the free
WPs at the same energy (dash-dotted curves) for $pp$ system at three
center of mass energies: $E_{\rm c.m.}=0.03$ MeV (a), $E_{\rm
c.m.}=0.133$ MeV (b) and $E_{\rm c.m.}=1.474$ MeV (c). }
\end{figure}

This statement is illustrated in Fig.~\ref{coul_fun} where the exact
Coulomb WPs for $pp$ system are compared to respective Coulomb
pseudostates found via diagonalization of the Coulomb Hamiltonian on
the free WP basis, and free WPs themselves at  the same energy. It
is clearly seen from the Figure the Coulomb WP $|x_i^C\rangle$ can
be  very accurately approximated by free WPs.

\section{Discrete version of the scattering theory}
The explicit relation between continuum wave functions and their WPs
makes it possible to develop a closed wave-packet formalism to treat
scattering in a very convenient discrete representation. The
complete formalism has been described in detail in our previous
papers~\cite{K2,KPR1} so that  we present here for the reader
convenience only the extract of the basic results.

\subsection{The wave packet space}
Let us call the linear shell spanned on basis of free WPs
$\{|x_i\rangle_{i=1}^{N}\}$ as an eigen wave-packet  space of the
Hamiltonian $h_0$. The projection operator onto this space
$\mathfrak{p}$ is defined in eq.~(\ref{projector}).

The property (\ref{projk}) allows one to find a finite-dimensional
representation for any operator $A=R(h_0)$ which depends on
Hamiltonian $h_0$ or just commute with it:
\begin{equation}
\mathfrak{A}\equiv
\mathfrak{p}A\mathfrak{p}=\sum_{i=1}^{N}|x_i\rangle A_i \langle
x_i|,
\end{equation}
where corresponding eigenvalues $A_i$ are given in an explicit form:
\begin{equation}
\label{ri} A_i=\frac1{B_i}\int_{\MD_i}dq R\left(
\frac{q^2}{2\mu}\right)|f(q)|^2.
\end{equation}
E.g., discrete eigenvalues of the free Hamiltonian are given via
eq.~(\ref{evh0}).

So that, using these properties one can build
finite-dimensional analogs for different scattering operators in the
free WP basis. In particular, the resolvent of the free Hamiltonian
 $g_0(E)=[E+i0-h_0]^{-1}$ is diagonal in the free WP representation:
\begin{equation}
\label{fdg0} \mathfrak{g}_0(E)\equiv
\mathfrak{p}g_0(E)\mathfrak{p}=\sum_{i=1}^{N} |x_i\rangle g_i(E)
\langle x_i|.
\end{equation}
Here the complex eigenvalues   $g_i(E)$ are determined according to
the general formula~(\ref{ri}).

For the charged particle scattering, one may easily build  the
finite-dimensional approximation for the Coulomb resolvent
$g_C=[E+i0-h_C]^{-1}$ in the Coulomb WP basis:
\begin{equation}
{\mathfrak g}_C(E)=\sum_{i=1}^N |x_i^C\rangle g_i(E)\langle x_i^C|
\end{equation}
The Coulomb resolvent eigenvalues
  $g_i(E)$ are found from the same eq.~(\ref{ri}) as for the free resolvent.
The explicit formulas  for these eigenvalues  are given in the
Appendix A.

The eigenvalues  $g_i(E)$ incorporate logarithmic singularities at
the bin endpoints. To smooth such undesirable singularities and to
convert our scheme to be completely discrete, one can employ an
additional averaging procedure over the ``on-shell'' energy bin to
represent the energy dependence of the resolvent.  As a result, we
get a purely finite-dimensional representation for the free
resolvent which is free of any singularities at real energies. In
other words, instead of the operator $\mathfrak{g}_0(E)$ which
continuously depends on energy one obtains the discrete set of
operators
\begin{equation}
\label{g0k}
\mathfrak{g}_0^k\equiv\frac1{D_k}\int_{\MD_k}\mathfrak{g}_0(E)dE,\quad
k=1,\ldots,N,
\end{equation}
 each of which is the averaged resolvent operator on the ``on-shell'' energy bin
$\MD_k$ where $D_k$ is its energy width.

\subsection{Correspondence between ``continuous'' and ``discrete''
quantities} Within the WP-formalism the discretization procedure
involving three steps is introduced:
\begin{itemize}
\item[(i)] Division of continuous spectrum of the free Hamiltonian onto non-overlapping intervals
and introduction of free WPs.
\item[(ii)] Projection of the scattering (and also bound-state )
 wave functions and operators onto the above  WP space.
\item[(iii)] An additional energy-averaging procedure for energy dependent operators.
\end{itemize}
It would be very useful to demonstrate how this discretization
procedure works in practical calculations by the example of solving
the Lippmann--Schwinger equation for the transition operator $t(E)$:
\begin{equation}
t(E)=v(E)+vg_0(E)t(E).
\end{equation}
After application of the above three steps (i)-(iii) one gets a
discrete set of operators
 $\mathfrak{t}^k$ at  $E\in\MD_k$ (instead of continuous operator $t(E)$). The matrix elements of
$\mathfrak{t}^k$ in the WP basis are related directly to
off-shell elements of $t$-matrix:
 \begin{equation}
 t(q,q',E)\approx\frac{[\mathfrak{t}^k]_{ij}}{\sqrt{D_i D_j}},\quad
  \left(\begin{array}{c}
          q\in\MD_i\\
          q'\in\MD_j\\
          E\in\MD_k\\
         \end{array}\right).
 \end{equation}
 These operators $\mathfrak{t}^k$ satisfy to simple matrix equations
 \begin{equation}
 \label{tk}
 \mathfrak{t}^k=\mathfrak{v}+\mathfrak{v}\mathfrak{g}_0^k\mathfrak{t}^k,\quad
 E\in\MD_k
 \end{equation}
where we denote by Gothic letters the WP projections of the
respective operators. From eq.~(\ref{tk}), one can get any on- and
off-shell $t$-matrix elements whose energy and momentum dependencies
are represented by histograms. It should be emphasized that
$t$-matrix constructed in the WP representation satisfies exactly
the unitarity relation \cite{KPR1}.

 Let's note that to find an elastic amplitude at
energy $E\in \MD_k$ one has to solve eq.~(\ref{tk}) for one column
of $t$-matrix  only, i.e. $[\mathfrak{t}^k]_{ik},i=1,\ldots,N$.

Finally, the $S$-matrix (and partial phase shift) can be found from
the relation:
\begin{equation}
S(E)\approx 1-2\pi i \frac{[\mathfrak{t}^k]_{kk}}{D_k},\quad
E\in\MD_k,
\end{equation}
where $D_k$ is the bin energy width.

By similar derivation one can build the WP analogs for the M\"oller
wave operators, total Hamiltonian resolvent, etc. \cite{KPR1}.

The Table~\ref{tab1} shows the one-to-one correspondence between
``discrete'' and initial ``continuous'' quantities.

\begin{figure}[h!]
\centering \epsfig{file=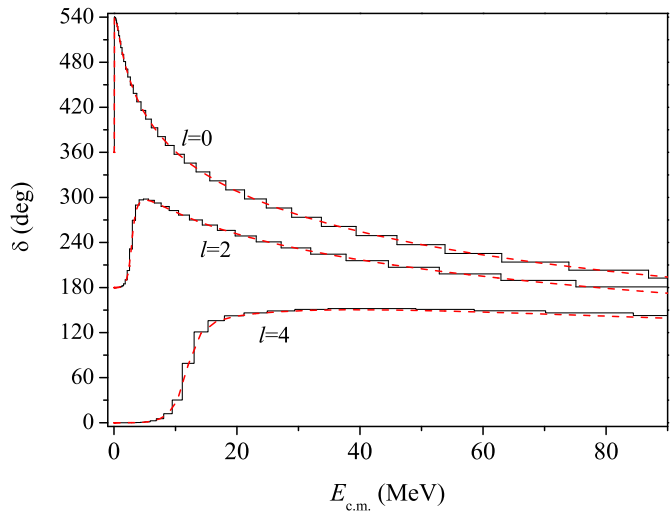,width=0.8\columnwidth}
\caption{\label{al_al} $s$-, $d$- and $g$-partial phase shifts of
 $\alpha-\alpha$ scattering found via the WP approach (solid curve) and
 from the direct solution of the Schroedinger equation
  by the Numerov method (dashed curve). }
\end{figure}

\begin{table}[h!]
\centering \caption{Comparison between the  continuous
representation for the scattering theory basic objects (in the
momentum space) and their discrete analogs in the WP subspace for
$E\in \mathfrak{D}_k$, $q\in\mathfrak{D}_i$ and
$q'\in\mathfrak{D}_{i'}$} \label{tab1}
\begin{tabular}{lccc}
\hline\noalign{\smallskip}
 & Continuous &\hspace{1cm} & Discrete WP   \\
\noalign{\smallskip}\hline\noalign{\smallskip}
1. The free resolvent & $g_0(E;q,q')$ & & $[\mathfrak{g}_0]^k_i \delta_{i,i'} .$\\
\noalign{\smallskip}\hline\noalign{\smallskip}
2. The total resolvent & $g(E;q,q')$ &  & $\mathfrak{g}_{i,i'}^k $ .\\
\noalign{\smallskip}\hline\noalign{\smallskip}
3. The $t$-matrix & $t(E; q,q')$ &  & $\mathfrak{t}_{i,i'}^k$.\\
\noalign{\smallskip}\hline\noalign{\smallskip}
4. The partial phase shift & $\delta(E)$& & $\delta^k$.\\
\noalign{\smallskip}\hline
\end{tabular}
\end{table}
As a good illustration of such a fully discrete technique for finding
the transition operator  we present here the solution of $\alpha$-$\alpha$
scattering problem where interaction includes both nuclear and
Coulomb potentials~\cite{KC}. The basic $s$-, $d$- and $g$- partial
$\alpha$-$\alpha$ phase shifts found using the above WP-technique
with additional energy averaging are displayed in Fig.~\ref{al_al}.

\section{Multichannel scattering problem}
Very often many-body scattering problem in atomic, nuclear and
molecular physics can be reduced to a multichannel scattering
although the number of channels to be incorporated may be very
large. Also a multichannel scattering problem arises in a simple
two-body case when an interaction potential is not spherically
symmetrical.

The total multichannel Hamiltonian of the system  can be written in
a matrix form (in the channel indices $\nu,\nu'$)
\begin{equation}
h_{\nu\nu'}=h_{0\nu}\delta_{\nu\nu'}+v_{\nu\nu'},\quad
\nu,\nu'=1,\ldots,K, \label{Rubtsova_Hamk}
\end{equation}
where $h_{0\nu}$ is the free-motion Hamiltonian in the channel $\nu$
with simple continuous spectrum $[\Omega_\nu,\infty)$, $\Omega_\nu$
is the channel threshold, and $v_{\nu\nu'}$ are the coupling
potentials. Further we will denote the operator matrices by the bold
characters. The matrix of the free multichannel Hamiltonian ${\bf h}_0$
is diagonal and therefore the continuous spectrum of ${\bf h}_0$ is the
union of spectra of $h_{0\nu}$. Thus, the continuous spectrum
of the multichannel Hamiltonian is degenerate (in contrast to a
single-channel case), and the multiplicity of the degeneracy $\eta$
being equal to the number of open channels and hence depends upon
energy. This means that at each energy there are $\eta$ scattering
wave functions corresponding to different boundary conditions.

\subsection{Free  WP basis for a multichannel scattering}
Now one has to define  free WP sets  for each free motion channel
Hamiltonian $h_{0\nu}$. We found that it is very convenient to
make the bin partitions for continuous spectra in various channels
so as the bins in different channels would be coinciding. In this
case, the interval endpoints  will be equal to the same energy
values for all open channels. By this  way one gets a degenerated
discretized  spectrum of the multichannel free Hamiltonian ${\bf
h}_0$. Such a discretization method, as will be demonstrated
below,  makes it possible to determine multichannel $t$-matrix
even without solving any scattering equations using properties of
multichannel pseudostates  \cite{KPRF}.

\begin{figure}[h!]
\centering \epsfig{file=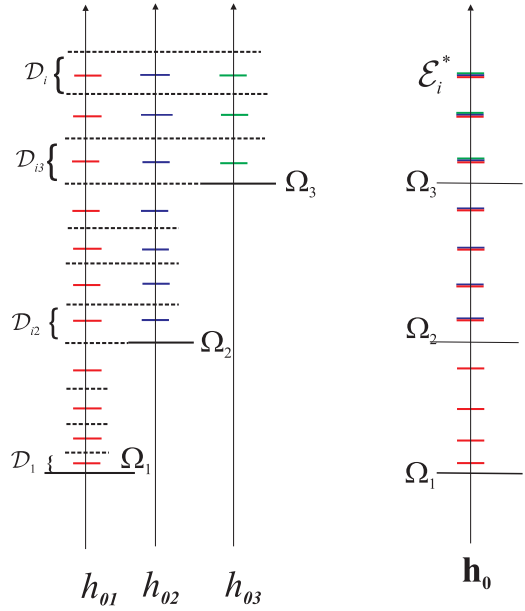,width=0.5\columnwidth}
\caption{\label{multi_free} The structure of the discretized
spectra of the separate free motion Hamiltonians  $h_{0\nu},
\nu=1,2,3,$ (on the left) and the resulting degenerate dicretized
spectrum of the total free Hamiltonian $\bf h_0$ (on the right).
The dashed lines represent bin boundaries while the solid lines
inside bins correspond to free WP eigenvalues $\ce_i^*$. }
\end{figure}
 So that, in the multichannel problem, one makes  a
  partition of the total
continuum as usual: the region $[\Omega_1,E_{\rm max}]$ is
 divided into
finite number of energy bins
$\MD_i\equiv[\ce_{i-1},\ce_{i}]_{i=1}^{N}$. Hereby an energy
threshold $\Omega_\nu$ should coincide with left end of some bin
which we denote as  $i_\nu$. Thus, for bins with indices $i<i_\nu$
there exist only  $\nu-1$ open channels. We assume that channels
are enumerated in the
 order of increasing  their threshold energies $\Omega_\nu$ and  $i_1=1$.
 Figure~\ref{multi_free}
demonstrates how the dicretized three-channel spectrum is
constructed.

   Using such partitions, the
corresponding set of free WPs in each initial channel $\nu$:
\begin{equation}
\label{Rubtsova_mxi}
 |x_{i}^{\nu}\rangle=\frac1{\sqrt{D_{i}}}
\int_{\MD_i}dE|\psi_{0}^{\nu}(E)\rangle,\ \nu=1,\ldots,K,\
i=i_\nu,\ldots,N
\end{equation}
can be constructed from the exact wave functions
$|\psi_{0}^{\nu}(E)\rangle$ of the free Hamiltonian $h_{0\nu}$.

After the introduction of multichannel WPs it is straightforward to
build a finite dimensional representation for  all  channel
resolvents  $g_0^{\nu}(E)=[E+i0-h_{0\nu}]^{-1}$ completely
analogously with a single-channel case. The channel resolvents
$g_0^{\nu}$ enter the expansion for the  multichannel free
resolvent. So that, this operator has the following
finite-dimensional approximation in the WP basis:
\begin{equation}
\mathfrak{g}_0(E)=\sum_{\nu=1}^K \sum_{i=i_\nu}^N|x_i^{\nu}\rangle
g_i^\nu(E) \langle x_i^{\nu}|,
\end{equation}
where complex eigenvalues are defined by the formulas similar to the
single-channel ones.

 Further one can project the basic scattering operators onto
such multichannel WP-basis similarly to the single-channel case
treated above. In particular, the transition matrix elements for
multichannel scattering problem is defined from the multichannel
Lippmann--Schwinger equation:
\begin{equation}
 t_{\nu\nu'}(E)=v_{\nu\nu'}+\sum_{\mu=1}^K v_{\nu\mu} g_0^\mu (E) t_{\mu\nu'}(E),\quad
 \nu,\nu'=1,\ldots,K.
\end{equation}
 So that, a solution for the multichannel scattering problem can be found
 by solving the matrix WP analog of the above equation which
is a direct generalization of the single-channel equation
(\ref{tk}).

However, there is another way by which one can solve the
multichannel problem in the WP representation using properties of
multichannel pseudostates.

\subsection{Eigenchannel representation and  a multichannel resolvent}
To study the resolvent of the total multichannel Hamiltonian $\bf
h$, we shall employ the multichannel formalism in the so called
Eigenchannel Representation (ER). The ER is the representation in
which the multichannel $S$-matrix takes a diagonal form with respect
to the channel indices \cite{greiner}. More definitely, one can
define at each energy $E$ two orthogonal sets of the scattering
functions for the Hamiltonian $\bf h$:
\begin{itemize}
\item[-] wave functions
$\{|\psi^{\nu}(E)\rangle\}_{\nu=1}^{\eta}$ corresponding to the
incoming waves in the channel $\nu$ (the so called experimental
channel representation) and \item[-] wave-functions
$\{|\tilde{\psi}^{\varkappa}(E)\rangle\}_{\varkappa=1}^{\eta}$
defined in the eigenchannel representation.
\end{itemize}
The ER states  differ from the experimental channel states by a
rotation in the channel space with
 the rotation matrix dependent on the energy $E$ \cite{KPRF,greiner}.

Just using the above ER formalism one can define
multichannel {\em scattering} wave-packets and derive an analytical
finite-dimensional representation for the total multichannel
resolvent (similarly to a one-channel case).

It is convenient to construct   multichannel scattering WP's as
integrals of the exact scattering wave functions for the total
Hamiltonian $\bf h$ defined in the ER:
\begin{equation}
\label{mult_zi} |z^{\vak}_k\rangle=
\frac{1}{\sqrt{C_{k}^{\vak}}}\int_{\De_k^{\vak}}
w(E)|\tilde{\psi}^{\vak}(E)\rangle dE,\quad
k=k_{\vak},\ldots,N^{\vak},
\end{equation}
where $\De_i^{\vak}\equiv[\ep_{i-1}^{\vak},\ep_i^{\vak}]$ are new
set of the total Hamiltonian bins whose parameters might be
different for different $\vak$.

Now we have to derive  a  spectral expansion of the total
multi-channel resolvent using multi-channel scattering states
defined in the ER. The spectral expansion of the resolvent in
the ``ordinary'' set of the initial  eigenfunctions of $h$ can be
written as sum of bound-state and continuum parts of the total
resolvent, viz. $g(E)=g^{\rm B}(E)+g^{\rm C}(E)$, where
\begin{equation}
 g^{\rm B}(E)=\sum_{n_b=1}^{N_b} \frac{|\psi_{n_b}\rangle
\langle\psi_{n_b}|}{E-E_{n_b}},\ g^{\rm
C}(E)=\sum_{\nu=1}^K\int_{\Omega_\nu}^{\infty}dE'
\frac{|\psi^{\nu}(E')\rangle\langle\psi^{\nu}(E')|}{E+{\rm i}0 -
E'}.
\end{equation}
 It is easy to show,
 that the following
relation is valid:
\begin{equation}
\sum_{\nu=1}^\eta|\psi^{\nu}(E')\rangle\langle\psi^{\nu}({E'})|=
\sum_{\varkappa=1}^\eta|\tilde{\psi}^{\varkappa}(E')\rangle\langle\tilde{\psi}^{\varkappa}(E')|,
\end{equation}
which corresponds to a rotation between experimental channel and
eigenchannel representations.

 Thus, we arrive at the spectral
expansion of the continuum part of the total resolvent in the
Hamiltonian eigenfunctions defined in ER:
\begin{equation}
g^{\rm C}(E)=\sum_{\varkappa=1}^K \int_{\Omega_\varkappa}^\infty
dE'\frac{|\tilde{\psi}^{\varkappa}(E')\rangle\langle\tilde{\psi}^{\varkappa}(E')|}{E+{\rm
i}0 - E'},
\end{equation}
where  the thresholds $\Omega_\vak$ in the ER  coincide with
thresholds of the initial experimental channels $\nu$.

Applying the  projection onto the multichannel basis
(\ref{mult_zi}) one gets the following finite-dimensional
representation for the total multichannel resolvent expressed via
the multichannel scattering WP basis
\begin{equation}
\label{Rubtsova_gcps} \mathfrak{g}^C(E)=
\sum_{\varkappa=1}^K\sum_{k=k_{\vak}}^{N} |z_{k}^{\varkappa}\rangle
g_{k}^\varkappa(E) \langle z_{k}^{\varkappa}| ,
\end{equation}
where complex eigenvalues $g_{k}^\varkappa(E)$ are defined as
integrals over discretization bins,
similar to  the single-channel ones.

 Now the question arises, how to
construct the states (\ref{mult_zi}) without solving the scattering
problem as we have done in a one-channel case. For this purpose we
employ a new treatment of multichannel pseudostates.

\subsection{Multichannel pseudostates and a solution of the multichannel scattering problem without
scattering equations} The continuous spectrum of the total
multichannel Hamiltonian $\bf h$ coincides with the spectrum of
${\bf h}_0$, so that at each energy there should exist $\eta$
independent solutions corresponding to different boundary conditions
for the scattering problem. Therefore, the usual $L_2$
discretization procedure becomes rather unclear here, because when
the total Hamiltonian is represented by a respective matrix in some
arbitrary finite $L_2$-basis the required multiplicity of the
spectrum disappears and one has only one pseudostate at every
discrete energy value which can hardly be treated properly. So, it
seems a one-to-one correspondence between the continuous spectrum
(which is multiply degenerated) and its discretized analog is lost
in the multichannel case.

\begin{figure}[h!]
\centering \epsfig{file=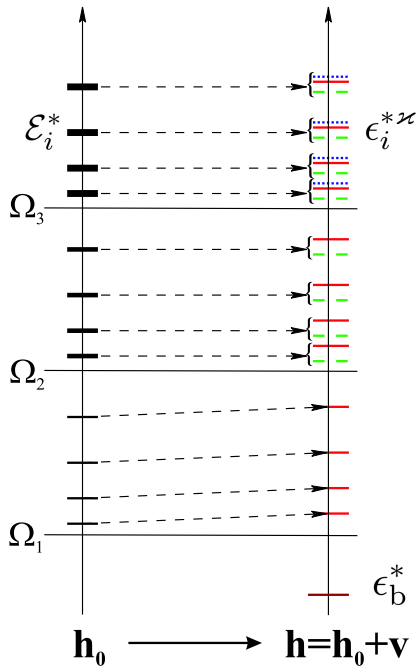,width=0.4\columnwidth}
\caption{\label{fig_spectrum} The  splitting of the eigenvalues in
the degenerated discretized spectrum of the multichannel Hamiltonian
$\bf h_0$
 caused by addition  the interaction
$\bf v$. $\Omega_\nu$ are the channel thresholds while  $\ep_b^*$
is the isolated eigenvalue corresponding to the single (in this
example) bound state of the total Hamiltonian $\bf h$. Sublevels
inside each splitted set corresponding to the same eigenchannel
number $\vak$ are shown by the same lines (i.e. solid, dashed and
dotted). }
\end{figure}
However,  the problem still can be solved if the discretized
 spectrum of the free multichannel  Hamiltonian
is   degenerate with necessary multiplicity as we constructed above.
As was shown in ref.~\cite{KPRF}, the application of the
perturbation $\bf v$
 to the unperturbed operator ${\bf h_{0}}$ with degenerate
 discretized spectrum  leads to a splitting of
each $\eta$-fold degenerate eigenvalue $\ce_i^{*}$ into the set of
$\eta$ perturbed eigenvalues
$\{\ep_i^{*\varkappa}\}_{\varkappa=1}^\eta$ of the total Hamiltonian
$\bf h$
 as is schematically
shown in Fig.~\ref{fig_spectrum}. The very important property of the
perturbed spectrum was shown in \cite{KPRF} (on the basis of results
from ref.~\cite{L1}), namely, these splitted sublevels form
different intermittent sets in the energy domain between thresholds
(i.e. the domain with fixed multiplicity $\eta$)
$\{\ep_i^{*\varkappa}\}_{i=i_{\eta}}^{i_{\eta+1}-1} $ each of which
corresponds to the $\varkappa$-th branch of the continuum spectrum.
So that the splitted levels are arranged in the same order (see
Fig.~\ref{fig_spectrum}):
 \[\ep_i^{*\varkappa +1}>\ep_i^{*\varkappa}, \quad i=i_{\eta},\ldots,i_{\eta+1}-1.\]
Moreover, this means that bands of splitted sublevels corresponding
to  different initial bins $i$
 are not overlapped with each other, which
 allows us to extract different $\vak$ branches from multichannel
 pseudostate spectrum.

Thus, one can select the new set of eigenvalues
$\{\ep_i^{*\varkappa}\}_{i=i_{\vak}}^{N^\vak}$ (and corresponding
functions of pseudostates) for each value $\varkappa$  which can
be treated as the $\varkappa$-th branch of the discretized
multichannel continuum  and
 be confronted with the  ER
scattering states  for respective  branch of the
continuous  spectrum of the total Hamiltonian~\cite{KPRF}.
 So, in this way, we can establish a  one-to-one
 correspondence between the discretized
spectrum of the multichannel Hamiltonian matrix and the continuous
spectrum of the initial multichannel Hamiltonian.  Due to such a
classification of branches in the discretized spectrum, one can
treat multichannel pseudostates on the same footing as  the
one-channel pseudostates. Now we can make the next  step ---
consider multichannel pseudostates  as approximations to
multichannel WPs defined in the ER (\ref{mult_zi}) (we have done the
same step in a one-channel case).

In this way, we  get  the multichannel scattering WP
 which are expressed from the free WPs by a simple
orthogonal transformation quite similarly to a  one-channel case:
\begin{equation}
\label{rotation_mult} |z^{\varkappa}_k\rangle \approx
\sum_{\nu=1}^\eta\sum_{i=i_\nu}^N O_{ki}^{\varkappa
\nu}|x_{i}^{\nu}\rangle.
\end{equation}
Now one can
construct the WP approximation for the total multichannel resolvent
via the relation (\ref{Rubtsova_gcps}) and get a final solution of
the multichannel scattering problem.

 As a good illustration of this approach  we have calculated the
coupled-channel $S$-matrix and the   partial phase shifts and the
mixing angle $\vep$ for the Nijmegen I  $NN$ potential \cite{nijm}
in coupled triplet $^3S_1-{^3}D_1$ channels  for the total spin
$S=1$ and isospin $I=0$. The two-channel $t$-matrix has been
obtained from the total resolvent using the well known formula
$${\bf t}(E)={\bf v}+{\bf v}{\bf g}(E){\bf v},$$ where finite-dimensional
 representation for the resolvent
(\ref{Rubtsova_gcps}) in the two-channel pseudostate basis is used.
More definitely, the multichannel $t$-matrix for $E\in\MD_i$ is
defined by the relation:
\begin{equation}
t_{\nu\nu'}(E)\approx \frac{\langle
x_i^\nu|v|x_i^{\nu'}\rangle}{D_i}+\frac{\langle x_i^\nu|v|z_{\rm
b}\rangle\langle z_{\rm b}|v|x_i^{\nu'}\rangle}{{D_i}(E-\ep_{\rm
b}^*)}
 +\sum_{\vak=1}^\eta
\sum_{k=k_{\vak}}^{N^\vak}\frac{\langle
x_i^\nu|v|z_k^{\vak}\rangle}{\sqrt{D_i}} g_{k}^\vak(E) \frac{\langle
z_k^\vak|v|x_i^{\nu'}\rangle}{\sqrt{D_i}},
\end{equation}
where $|z_{\rm b}\rangle$ is a WP approximation for the bound-state
wave function with energy $\ep_{\rm b}^*$ and  matrix elements of
the interaction potential $\bf v$ between free and scattering WP
states can be written using expansion (\ref{rotation_mult}) as
follows:
 \begin{equation}
\langle x_i^\nu|v|z_k^{\vak}\rangle=\sum_{\mu=1}^\eta
\sum_{j=j_{\mu}}^N O^{\vak\mu}_{kj}\langle
x_i^\nu|v|x_j^{\mu}\rangle.
 \end{equation}
 Thus, in the developed approach we can find the accurate multichannel $t$-matrix
{\em without solving any scattering equations at all}.

\begin{figure}
\centering \epsfig{file=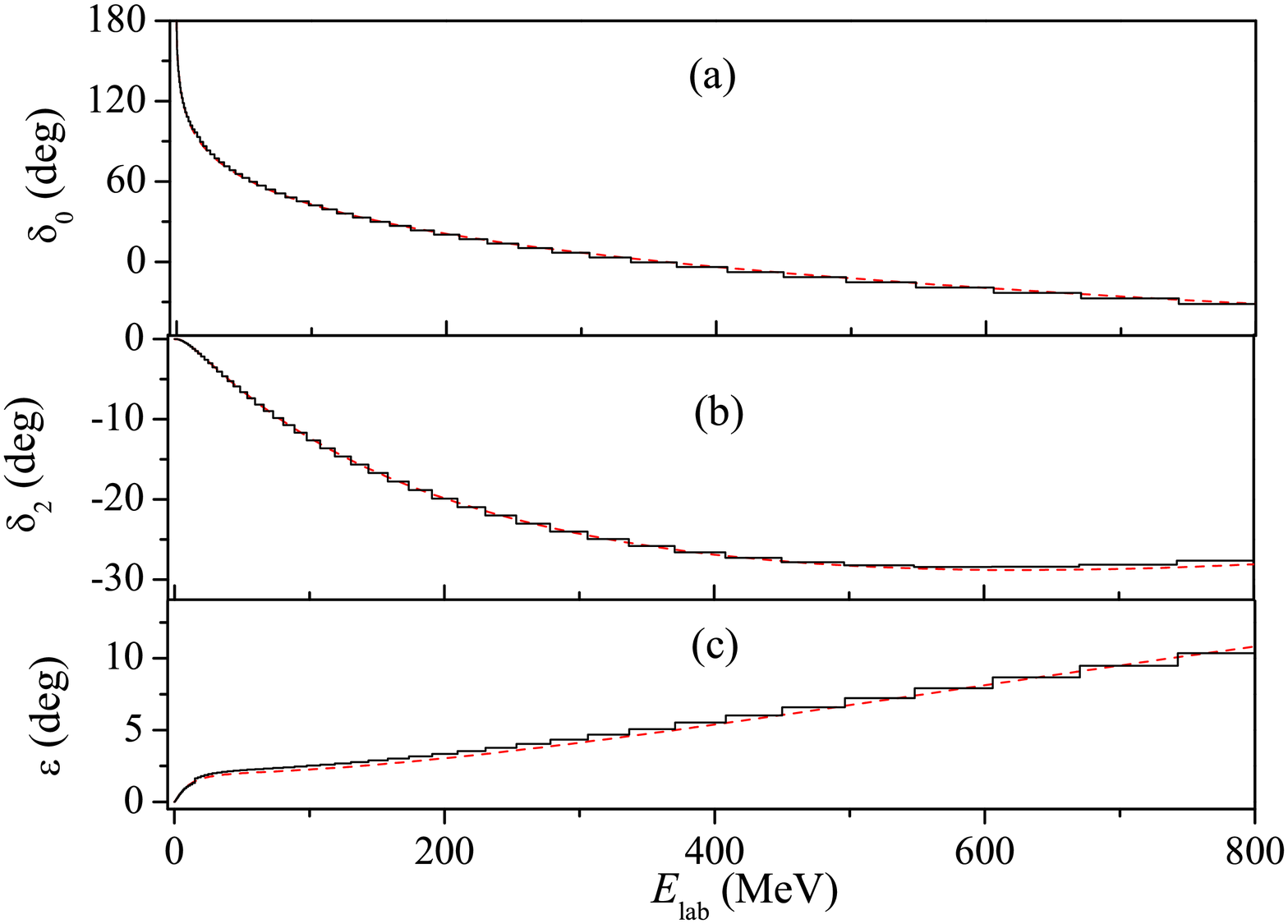,width=0.8\columnwidth}
\caption{\label{fig_multi} The partial phase shifts $\delta_0$ (a),
$\delta_2$ (b) and the mixing parameter $\vep$ (c) for the Nijmegen
$NN$ potential found from two-channel Hamiltonian diagonalization in
the WP basis (solid curves) and from the direct numerical solution
of the two-channel Lippmann--Schwinger equation (dashed curves).}
\end{figure}

In  Fig.~\ref{fig_multi} the partial phase shifts $\delta_0(E)$,
$\delta_2(E)$, and mixing parameter $\vep(E)$ (i.e. those which
enter the $NN$ differential scattering cross sections -- in the
Stapp parametrization) are shown in very wide energy range $0<E_{\rm
lab}< 800$ MeV. They are found from just a single diagonalization of
the respective two-channel Hamiltonian of the $NN$ interaction using
a two-channel free WP-basis.
  We compare these quantities  in  Figure to those obtained from direct
numerical solutions of the two-channel Lippmann--Schwinger
integral equation at many energies in the above energy range.

This comparison shows that  one can derive the
accurate multichannel $t$-matrix or total resolvent for general
multichannel case using a single diagonalization of the Hamiltonian matrix,
i.e. without solving any scattering equations. It opens a new way in complicated
coupled-channel calculations in atomic, molecular and nuclear physics.

\section{Stationary wave packets in a three-body problem}
\label{three} Let's consider now a general three-body scattering problem
for particles 1,2 and 3, interacting via pairwise short-range
potentials $v_a$ $(a=1,2,3)$. It is convenient to use here three
Jacobi momentum sets $({\bf p}_a,{\bf q}_a)$ corresponding to three
channel Hamiltonians $H_a$ $(a=1,2,3)$ which define
 the asymptotic states of the system.
The symbol $p_a$ denotes  the relative momentum of   $b$
and $c$ particles while  $q_a$ is the momentum of the third particle
relative to the c.m. of the pair
 $\{bc\}$. In a general
 case of different particles  the respective
wave-packet bases should be constructed independently for each
Jacobi set~\cite{K2,KPR1}. Below we show how to built one of those.

The  channel Hamiltonian (e.g. $H_1$ for $a=1$)
 takes the form  of a
direct sum of two sub-Hamiltonians
\begin{equation}
\label{ch_ham} H_1\equiv h_1\oplus h_{0}^1,
\end{equation}
where sub-Hamiltonian $h_1=h_0+v_1$ includes the interaction  $v_1$
in the $\{23\}$-subsystem  and the sub-Hamiltonian $h_0^1$
 corresponds to a free relative motion of the center of mass for particles 2 and 3
 and the spectator particle 1.

\subsection{The lattice three-body basis}
The three-body wave-packet basis   is constructed at first for the
three-body free Hamiltonian defined in the same Jacobi set:
\begin{equation}
H_0=h_0\oplus h_0^1.
\end{equation}
 So that the three-body free
wave-packet states are built as direct products of the respective two-body
wave-packet states. To construct the
three-body free WP basis functions, we introduce partitions of the continua of two free
 sub-Hamiltonians, $h_0$ and $h_0^1$, onto non-overlapping intervals
 $\{\mathfrak{D}_i\equiv[\ce_{i-1},\ce_i]\}_{i=1}^N$ and
 $\{\bar{\mathfrak{D}}_j\equiv[\bar\ce_{j-1},\bar\ce_j]\}_{j=1}^{\bar N}$
 respectively and introduce two-body free WPs as in eq.~(\ref{ip}):
\begin{equation} \label{pq}
|x_i\rangle=\frac{1}{\sqrt{B_i}}\int_{\mathfrak{D}_i}f(p)|\psi_{0p}\rangle
dp,\ |{\bar
x}_j\rangle=\frac{1}{\sqrt{\bar{B}_j}}\int_{\bar{\mathfrak{D}}_j}\bar{f}(q)|\psi_{0q}\rangle
dq,
\end{equation}
where $p$ and $q$ are Jacobi momenta and  $B_i,\bar{B}_j$ and
$f(p),\bar{f}(q)$ are normalization factors and weight functions
respectively. Here and below we denote
 functions and values corresponding to the variable $q$ with
additional bar mark to distinguish them from  functions
corresponding to the variable $p$.

 When constructing the three-body WP basis one should take
into account  spin and angular parts of the basis functions. Thus
the three-body basis function can be written as:
\begin{equation}
\label{xij} |X_{ij}^{\Ga\al\be}\rangle\equiv |x_i^\al,{\bar
x}_j^\be;\al,\be:\Ga\rangle=|x_i^\al\rangle\otimes|{\bar
x}_j^\be\rangle |\alpha,\beta:\Gamma\rangle,
\end{equation}
where $|\alpha\rangle$ is a spin-angular state for the $\{23\}$
pair, $|\beta\rangle$  is a  spin-angular state for the third
particle, and $|\Gamma\rangle$ is a set of  three-body quantum
numbers. The state (\ref{xij}) is the WP analog of the exact free
motion  state in three-body continuum
$|p,q;\alpha,\beta:\Gamma\rangle$ for the three-body free
Hamiltonian $H_0$ \cite{Gloeckle_rep}.

The  three-body free WP basis functions (\ref{xij}) are constant
inside the cells of the momentum lattice built from two
one-dimensional cells $\{\mathfrak{D}_i\}_{i=1}^{N}$ and
$\{\bar{\mathfrak{D}}_j\}_{j=1}^{\bar N}$ in momentum space. We
refer to {\em the free WP} basis as {\em a lattice} basis and
denote the respective two-dimensional bins (i.e. the lattice
cells) by $\MD_{ij}=\MD_i\otimes\BMD_j$.

Such multi-dimensional wave-packet basis is an ``eigenbasis''  for
the
 free Hamiltonian  $H_0$ and has the same properties with
respect to this operator as two-body free WPs with respect to the
two-body free Hamiltonian $h_0$. In particular, every operator which
functionally depends on   $H_0$ has an explicit  finite-dimensional
representation in the  lattice basis discussed \cite{KPR1}. So that,
it is straightforward to obtain {\em a diagonal representation} for
the three-body free resolvent  $G_0$ and to use it further for
solving few-body scattering equations. This is a way to develop the
wave-packet matrix scheme which is a finite-dimensional form of the
initial scattering equations.

However, the above discrete representation has a remarkable
advantage in comparison with the continuous one: one can construct
and employ a few-body wave-packet basis directly for the three-body
channel Hamiltonian $H_1$. This allows to simplify a solution of
scattering problems drastically.

\subsection{The WP basis for the channel three-body Hamiltonian}
  Let's  introduce a WP basis for
   two-body sub-Hamiltonian $h_1$.
 Assume as above that there are $N_{\rm b}$ bound states  in the $\{23\}$ subsystem with
 corresponding bound-state wave functions $\{|{z}_n^\alpha\rangle\}_{n=1}^{N_{\rm b}}$ and
 eigenenergies $\{\epsilon_n^{\alpha*}\}_{n=1}^{N_{\rm b}}$. One defines the
 partition $\{\Delta_k\}_{k=N_{\rm b}+1}^{N^\alpha}$ of the continuous spectrum
of $h_1$  and constructs the set of   scattering wave packets
$|z_k^\alpha\rangle$ from the respective exact
 scattering wave functions $|\psi_p^\alpha\rangle$ according to eq.~(\ref{zi}).
   The complete WP basis
$\{|z_{k}^\alpha\rangle\}_{k=1}^{N^\alpha}$ for the $h_1$
sub-Hamiltonian includes  bound state functions and  scattering wave
packets and its functions may depend on possible spin-angular
quantum numbers.

  Now one can build the  three-body wave-packets (3WP) for the
  channel Hamiltonian $H_1$
just as products of two types of WP states for $h_1$ and $h_0^1$
sub-Hamiltonians
 whose  spin-angular parts are combined to
the respecive three-body states with quantum numbers $\Ga$:
 \begin{equation}
\label{si} |Z^{\Ga\al\be}_{kj}\rangle
\equiv|z_k^\al,\bar{x}_j^\be,\al,\be:\Ga\rangle,\quad
{k=1,\ldots,N^\al},\quad j=1,\ldots,\bar{N}.
\end{equation}
 The properties of the
3WP constructed in this way are the same as properties of the
two-body wave packets  viz. they form an orthonormal set and any
operator functionally dependent on the channel Hamiltonian $H_1$
has a diagonal matrix representation in the subspace spanned on
this basis. It allows us to construct a finite-dimensional
approximation for the three-body channel resolvent $G_1(E)\equiv
[E+{\rm i}0-H_1]^{-1}$ in a very convenient analytical form.

 Indeed,  the exact three-body channel resolvent  is a convolution of the two-body
 subresolvents $g_1(E)=[E+i0-h_1]^{-1}$ and $g_0^1(E)=[E+i0-h_0^1]^{-1}$ \cite{K2,K3,K4,KPR1}:
\begin{equation}
G_1(E)=\frac{1}{2{\pi\rm i}}\int_{-\infty}^{\infty}{\rm d}\epsilon
g_1(\epsilon)g_0^1(E-\epsilon).
\end{equation}
Using spectral expansions for these two-body resolvents and making
the integration, one gets an explicit expression for the exact
channel resolvent $G_1$
 as a sum of two terms $G_1(E)=G_1^{\rm BC}(E)+G_1^{\rm
CC}$, where  the bound-continuum (BC) part  takes the form
\cite{KPR1}:
\begin{equation}
G_1^{\rm BC}(E)=\sum_{\Ga,\al,\be}\sum_{n=1}^{N_{\rm
b}}\int_{0}^\infty {\rm d}q
\frac{|z_n^\al,\psi_{0q}^\be;\al,\be:\Ga\rangle\langle
z_n^\al,\psi_{0q}^\be;\al,\be:\Ga|}{E+{\rm
i}0-\epsilon_n^{\al*}-\frac{q^2}{2M}},
\end{equation}
where $M$ is the reduced  mass in the $\{23\}+1$ channel.   While the
continuum-continuum (CC) part of $G_1$ takes the form:
\begin{equation}
G_1^{\rm CC}(E)=\sum_{\Ga,\al,\be}\int_{0}^\infty {\rm d}p
\int_{0}^\infty {\rm d}q
\frac{|\psi_p^\al,\psi_{0q}^\be,\al,\be:\Ga\rangle\langle
\psi_p^\al,\psi_{0q}^\be,\al,\be:\Ga|}{E+{\rm
i}0-\frac{p^2}{2\mu}-\frac{q^2}{2M}},
\end{equation}
where $\mu$ is the reduced mass in the $\{23\}$ subsystem.

Projecting further the exact channel resolvent onto the three-body
channel 3WP basis defined in eq.~(\ref{si}), one can find analytical
formulas for the matrix elements  of the $G_1$ operator. The
respective matrix is diagonal in all wave-packet indices:
\begin{equation}
\mathfrak{G}_1=\sum_{\Ga,\al,\be}\sum_{k,j}
|Z^{\Ga\al\beta}_{kj}\rangle G^{\Gamma\al\beta}_{kj}(E) \langle
Z_{kj}^{\Ga\al\beta}|.
 \label{dg1}
\end{equation}
The matrix elements $G^{\Gamma\al\be}_{kj}$ are defined  as
integrals over the respective momentum bins for the BC part:
\[
G^{\Gamma\al\be}_{kj}=\frac1{{\bar
B}_j}\int_{\BMD_j}\frac{|\bar{f}(q)|^2{\rm d}q }{E+{\rm
i}0-\epsilon_k^{\al*}-\frac{q^2}{2M}},\quad k=1,\ldots,N_{\rm
b},\quad j=1,\ldots,\bar{N}\quad \eqno(\ref{dg1}a)
\]
 and for the CC-one:
\[
G^{\Gamma\al\be}_{kj}=\frac1{C_k^{\al}
\bar{B}_j}\int_{\De_{k}^{\al}}\int_{\BMD_j}
\frac{|w(p)|^2|\bar{f}(q)|^2{\rm d}p{\rm d}q }{E+{\rm
i}0-\frac{p^2}{2\mu}-\frac{q^2}{2M}},\quad k=N_{\rm
b}+1,\ldots,N^{\al},\quad j=1,\ldots,\bar{N}.\quad \eqno(\ref{dg1}b)
\]
  These elements depend on the spectral partition parameters (i.e.
$\De_k^\al$ and $\BMD_j$ bin endpoints) and total energy $E$ only.
They
 do not depend explicitly on the interaction
potential $v_1$. When solving the scattering equations in the
finite-dimensional WP basis   the corresponding solution converges
with increasing the basis dimension,  the final result turns out to
be {\em independent} of the particular spectral partition
parameters. Explicit formulas for the resolvent matrix elements
(\ref{dg1}a) and (\ref{dg1}b) are given in Appendix A.

The explicit  analytical representation (\ref{dg1}) for the channel
three-body resolvent  is a basic feature for
  the wave-packet
approach since it allows us to simplify solution of the general
three-body scattering problem drastically. In particular, this
representation
 has been used
to solve the finite-dimensional analog for the Faddeev equations
~\cite{KPR2,KPR_br}. This representation has been  employed also to
solve some particular three-body scattering problems using the
three-body Lippmann--Schwinger equation \cite{K3,K4,Moro} for
composite projectile scattering off nuclear target (see below).

To use the states (\ref{si}) practically, one can approximate them
with the pseudostates of the sub-Hamiltonian $h_1$ in some $L_2$
basis. As  has been shown earlier in  Section 3, the free WP basis
is very appropriate to approximate scattering states because the
respective functions have a very long-range behavior in
configuration space which makes it possible to describe properly an
asymptotical behavior  of Faddeev wave function components at large
distances. So one can calculate the eigenstates (the bound and
pseudostates) of the sub-Hamiltonian $h_1$ matrix in the two-body
WP-basis $\{|x_i\rangle\}_{i=1}^N$ via a diagonalization procedure.
As a result one gets the eigenstates of the $h_1$ sub-Hamiltonian
expanded in the free WP basis (for each quantum  number $\al$)
similar to eq.~(\ref{rot1}).

Hence, starting from the free WP bases for each two-body
sub-Hamiltonian one gets a set of three-body basis states both for
free and channel Hamiltonians, $H_0$ and $H_1$ respectively, which
are related to each other by a simple matrix rotation. Using
projection of operators and wave functions onto these 3WP bases it
is possible to solve three-body scattering problems via matrix
formalism.

It should be mentioned that in a general few-body case, the free and
the channel wave-packet bases can be constructed by a very similar
procedure using a continuum discretization for every  subsystem. In
the bases constructed in this way, it is straightforward  to find
explicit finite-dimensional representations for channel and free
resolvents and furthermore to solve the resulting scattering
equations in a very convenient matrix form.

\section{Composite particle scattering off a nuclear target}
Let us consider a  few-body  problem  of an elastic composite
particle scattering off a target when rearrangement channels are
neglected while  intermediate excitations into continuum are taken
into account.

\subsection{Solution of the  Lippmann--Schwinger equation}
The total Hamiltonian for the composite projectile plus target
system can be written as follows:
\begin{equation}
\label{ham_proj} H=h_{\rm int}+h_C+V_{\rm ext},
\end{equation}
 where $h_{\rm
int}=h_0+\displaystyle{\sum_{i<j}}v_{ij}(r_{ij})$ is the internal
sub-Hamiltonian for the projectile composed from several fragments
$i=1,\ldots,K$,
 while  $h_C (R)$ is the
 sub-Hamiltonian of projectile asymptotic motion  including  the center of mass Coulomb
interaction, and
  $V_{\rm ext}=\displaystyle{\sum_{i}} v_i(r_i)$ are the fragment-target external interactions where
  $r_i$ are separate fragment positions and $r_{ij}$ are their relative distances.

 The optical potentials $v_{i}$ are usually assumed to be complex and
energy-dependent. However,  the smooth energy dependence of the
input optical potential is of no significance  for our present
purpose and, thereby we omit it here. Also we assume  that there are
bound states  in $h_{\rm int}$ sub-Hamiltonian
$\{|z_n\rangle\}_{n=1}^{N_b}$  with energies
$\{\epsilon_n^*\}_{n=1}^{N_b}$.

The channel Hamiltonian is a direct sum of two sub-Hamiltonians:
\begin{equation}
\label{hch} H_{\rm ch}=h_{\rm int}\oplus h_C.
\end{equation}
In the  case to be discussed, when  rearrangement channels are
neglected, the transition operator $T$ which describes the elastic
scattering of the  projectile by the target nucleus as well as the
projectile breakup is found from a single Lippmann--Schwinger
equation (LSE) \cite{Moro}:
  \begin{equation}
  \label{LS}
  T(E)={V}_{\rm ext}+{V}_{\rm ext}G_{\rm ch}(E)T(E),
  \end{equation}
where $G_{\rm ch}=[E+i0-H_{\rm ch}]^{-1}$ is the channel resolvent
which determines the asymptotic states in the elastic channel.

Then, wave-packet basis states corresponding to a discretization of
the channel Hamiltonian continuum are built as products of $h_{\rm
int}$ and $h_C$ sub-Hamiltonian WP basis states as defined in
eq.~(\ref{si}). The only difference in that we have to replace free
WPs over the center of mass variable with Coulomb WPs for the
sub-Hamiltonian $h_C$. Thus our  WP basis in this case takes the
form:
\begin{equation}
\label{zi1}
|Z_{kj}^{Ll\la}\rangle=|z_k^l,\bar{x}_j^{C\la};l,\la:L\rangle,
\end{equation}
where $l$ and $\la$ are subsystem orbital momenta while $L$ is a
total orbital momenta of the system which assumed to be conserved.

Thus, after a WP-projection of   wave functions and scattering
operators one gets  respective vectors and matrices in the channel
WP space. Such a wave-packet representation is
 eigen for the projected
channel Hamiltonian  and the basis states from eq.~(\ref{zi1})
correspond to exact asymptotic states of the system. In particular,
the initial state wave function of the system which defines the
asymptotic free motion of the projectile corresponds
 to a single WP-state. For example, if one study an elastic scattering of the projectile
 in its ground state (assume that  $l=0$), the initial state wave
 function
$|\Psi_0^L(E)\rangle\equiv|z_1^0,\psi^L(E-\epsilon_1^*);L\rangle$ is
represented by the state $|Z^L_{1j_0}\rangle$ of the channel WP
basis  where $E-\epsilon_1^*\in\BMD_{j_0}$.

Eventually, after   the wave-packet projecting of the scattering
operators,
 all  terms in the  LSE (\ref{LS}) are reduced to a matrix form
and thus the $T$-matrix can be found  from a single matrix equation:
\begin{equation}
\label{bold_t} \mathbb{T}=\mathbb{V}+\mathbb{V}\mathbb{G}_{\rm
ch}\mathbb{T},
\end{equation}
where $\mathbb{V}$ is the external interaction matrix in the
three-body wave-packet basis and  $\mathbb{ G}_{\rm ch}$ is a
diagonal matrix of the channel resolvent (\ref{dg1}) taken at the
total energy
 $E$.
It should be stressed that to find   on-shell and half-shell
$T$-matrix elements, it is sufficient to solve the matrix equation
(\ref{bold_t}) only for one column  $t_{n}\equiv T_{n,n_0},\quad
n=1,\ldots,N\cdot \bar{N}$ (we use here the multi-index $n=(k,j)$,
so that the label $n_0=(1,j_0)$ corresponds to the initial state).
The respective solution can be found from the equation
\begin{equation}
\label{t_lin} ({\mathbb I}-{\mathbb V}{\mathbb G}_{ch})t=v.
\end{equation}
Here the notation is: $v_n\equiv V_{n,n_0},\quad
n=1,\ldots,N\cdot\bar{N}$, and $\mathbb I$ is the unit matrix.
Finally, the $S$-matrix elements are interrelated to the vector of
$t$-matrix elements by the relationship \cite{Moro}:
\begin{equation}
\label{st} S_{\rm el}(E)=1-2\pi i \frac
{t_{n_0}}{\bar{D}_{j_0}},\quad (E-\epsilon_1^*)\in\BMD_{j_0}.
\end{equation}

\begin{figure}
\centering\includegraphics[width=8.5cm]{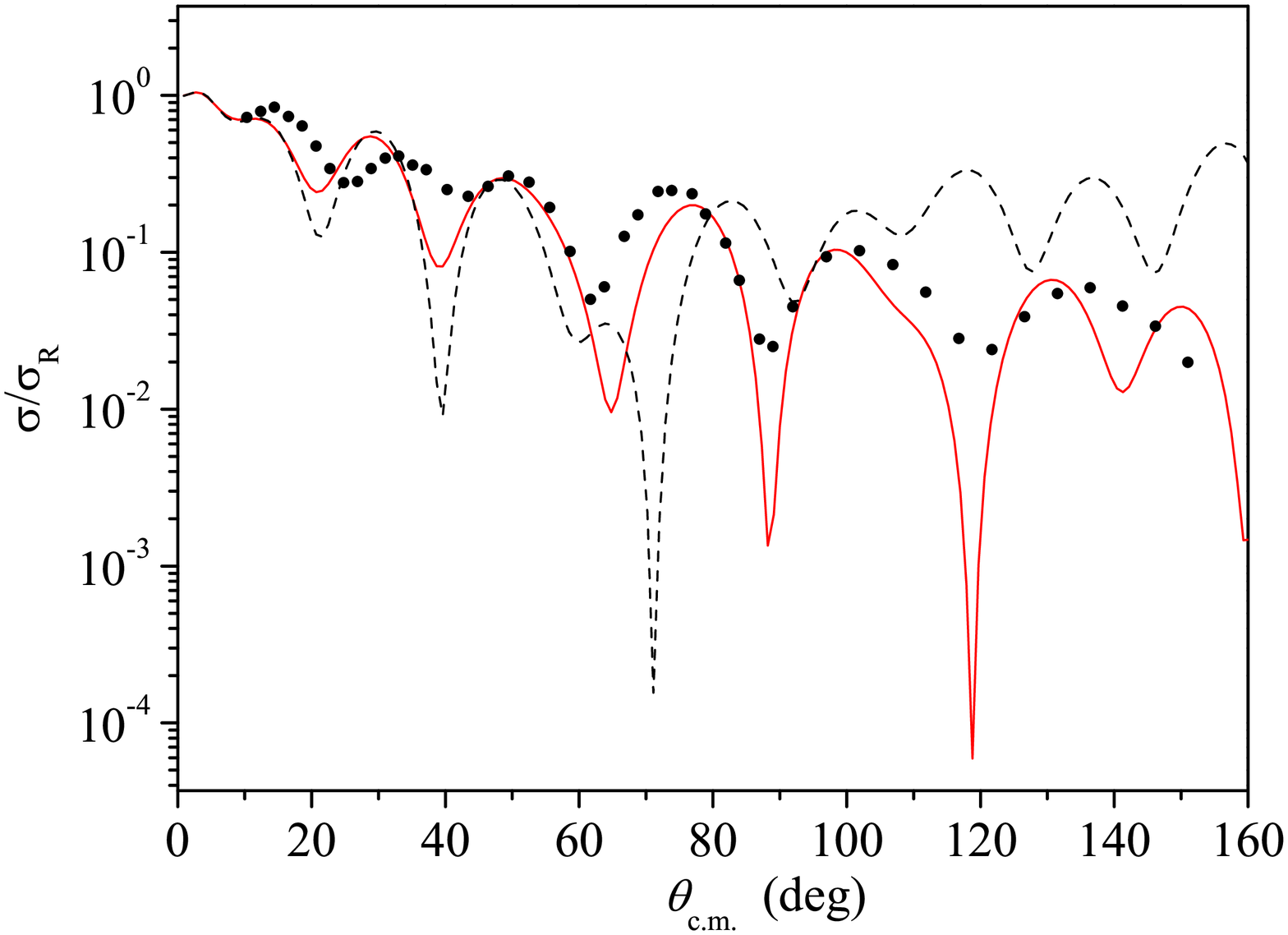} \caption{\label{fig4}
The comparison of the deuteron off $^{58}{\rm Ni}$-target elastic
cross sections (in ratio to the Rutherford cross section
$\sigma_R(\theta)$) obtained in the WPCD approach with an inclusion
of intermediate deuteron breakup channels (full curve), with the
folding model, i.e. without taking into account intermediate breakup
channels, (dashed curve) and the experimental data (circles)
\cite{expNi}.}
\end{figure}

As a particular example for such a scattering problem we consider
the well known test case: the deuteron elastic scattering off
$^{58}$Ni target   --- see the respective results in
Fig.~\ref{fig4}. The details of a calculation and  potential
parameters used can be found in ref.~\cite{K4}. Also we have shown
\cite{Moro} that our discrete WP-solution for such a coupled-channel
problem is in a very good agreement with that derived from the
traditional CDCC approach in which one has to solve a large system
of coupled differential equations in every partial wave.

\subsection{Construction of an effective projectile-target interaction}
In many problems of quantum physics it is necessary to know an
operator of effective interaction between a composite projectile and
stable target or vice versa, which takes into account  the virtual
excitation of the incident particle (or target), i.e. its dynamic
polarization in the scattering process. Such an operator of the
effective interaction is replaced usually by some phenomenological
optical potential between composite projectile and stable target (or
vise versa) because its microscopic evaluation is often a very
cumbersome problem. So the convenient approach for a theoretical
calculation of the effective operator is a very good object for the
theory.

 The wave-packet discretization method allows us to construct such
operators explicitly, using the finite-dimensional representation of
the channel resolvent obtained above. In this solution it is
convenient to apply the Feshbach projection operator formalism
\cite{feshbach}.

For the reader's convenience, we briefly recall a derivation of the
effective operator of interaction within such projection approach
for the case of composite particle scattering off a nucleus
discussed in the previous subsection.
  Consider the Schroedinger equation
 $$
 H|\Psi\rangle = E|\Psi\rangle
 $$ for the wave function of the system defined by the Hamiltonian (\ref{ham_proj}).

Let, $|z_1\rangle$ be the projectile ground state with energy
$\epsilon_1^*$.
 Using
the above projection operator technique, one  defines two
projection operators:
\begin{equation}
 F=|z_1\rangle\langle z_1|,\quad Q=1-F,\quad QF=0.
\end{equation}
Here $F$ is the projector onto the elastic channel, $Q$ is the orthogonal
projector onto all the inelastic channels. As a result the Schroedinger equation for
the total wave function is splitted into two coupled equations for components
 $F|\Psi\rangle$ and $Q|\Psi\rangle$ :
\begin{eqnarray}
(FHF - E)F|\Psi\rangle = - FHQ|\Psi\rangle,\nonumber\\
(QHQ - E)Q|\Psi\rangle = - QHF|\Psi\rangle .\label{pqpsi}
\end{eqnarray}
Further, we introduce the resolvent of the Hamiltonian
projected onto  $Q$-subspace: $G_Q^{(+)}(E)=[(E+i0)-QHQ]^{-1}$ .
Then, substituting the second equation in (\ref{pqpsi}) to the first one we obtain the
well known equation for the elastic scattering function $F|\Psi\rangle$:
\begin{equation}
\label{maineq} (FHF+FU_{\rm eff}(E)F-E)F|\Psi\rangle =0,
\end{equation}
where
\begin{equation}
\label{ueff} FU_{\rm eff}(E)F\equiv FHQG_Q^{(+)}QHF
\end{equation}
is an effective nonlocal and energy-dependent  operator of interaction between
a composite particle
and the target with taking into account all the inelastic channels.
  Since the operators $F$
and $h_{\rm int}\oplus h_C$ commute, and the orthogonality condition
$FQ=0$ is valid, the equation (\ref{maineq})
  takes the form of a conventional two-particle Schroedinger equation
   for projection $|\chi_1\rangle=\langle z_1|\Psi\rangle$ of the total wave function onto the elastic
   channel:
\begin{equation}
[h_C+V_{\rm fold}+U_{\rm eff}-(E-\ep_1^*)]|\chi_1\rangle=0.
\end{equation}
Here we have introduced a folding potential $V_{\rm fold}$
describing the interaction of the projectile center of mass  with
the target nucleus when the inelastic channels corresponding to the
excitation or breakup of the incident composite particle are fully
neglected:
\begin{equation}
\label{vatanab} V_{\rm fold }(R)=\langle z_1|V_{\rm ext}| z_1\rangle.
\end{equation}
Here the integration in the
 the matrix element is carried out only over the internal variables.
 All inelastic (and breakup) effects are included into the
non-local Feshbach potential:
\begin{equation}
\label{pot} U_{\rm eff}(E)=\langle z_1|V_{\rm ext}Q G_Q(E)QV_{\rm
ext}|z_1\rangle.
\end{equation}
Despite the compact notation, the computation of this operator is in
principle even more difficult problem than the solution of the
original scattering problem. However, when using the wave-packet basis (\ref{zi1})
consisted from eigenfunctions of the channel Hamiltonian the practical
construction of operator $U_{\rm eff}(E)$  becomes clear and
straightforward.

Indeed, using
  the wave-packet
states for the channel Hamiltonian, one finds the
finite-dimensional diagonal representation for the channel
resolvent $G_{\rm ch}(E)$ obtained via formulas from eq.~(\ref{dg1}).

In the WP-approach the $Q$-subspace is just orthogonal part of the
WP subspace, so that the finite-dimensional representation for the
projector $Q$ can be written as (we omitted possible spin-angular
indices):
\begin{equation}
\mathfrak{Q}={\sum_{k,j}}'|z_k,\bar{x}_j^C\rangle\langle
z_k,\bar{x}_j^C|,
\end{equation}
where upper prime symbol means that the sum does not include the
ground state of $h_{\rm int}$.
 Thus  the matrix
representation for the projected $G_Q$ operator is straightforward:
\begin{equation}
\label{qgq}
{\mathfrak{G}}_Q(E)=\{[{\mathfrak {G}}_{\rm
ch}^Q(E)]^{-1}-{\mathfrak {V}}^Q_{\rm ext}\}^{-1},
\end{equation}
 where all operators are meant as being projected onto the
$Q$-subspace.

 After direct evaluation of the matrix for $G_Q$-operator from eq.~(\ref{qgq}),
 the explicit formula
for the  effective potential in the wave-packet representation can
be written in the form:
\begin{equation}
U_{\rm eff}(E,R,R')={\sum_{kj}}'{\sum_{k'j'}}' B_{kj}(R) [{\mathbb
G}_Q(E)]_{kj,k'j'}B_{k'j'}^{*}( R'),
\end{equation}
 where the
form-factors are defined as the integrals:
\begin{equation}
 B_{kj}(R)=\langle z_1|V_{\rm
ext}|z_k,\bar{x}_j^C\rangle,
\end{equation}
 and an integration is done over internal
variables only. In practical calculations, all the necessary
spin-angular parts of wave functions should be taken into account.

Thus, the WP approach gives a direct and convenient way to calculate
an effective optical potential for an interaction of a composite
particle with a stable target, also this  formalism can be
generalized for constructing an effective interaction between two
colliding composite particles which may be excited or disintegrate
in the scattering process.

\begin{figure}
 \centering \epsfig{file=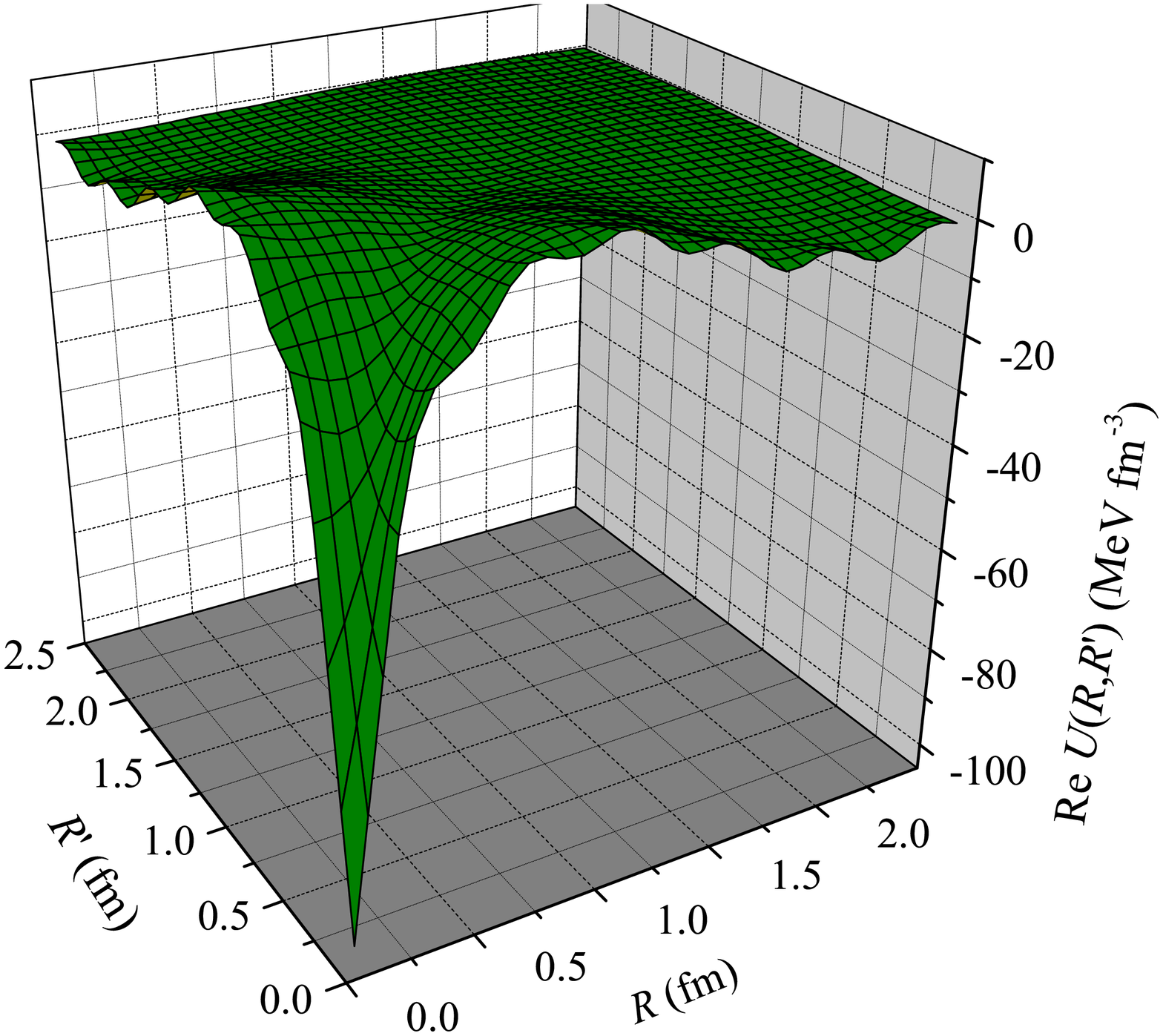,width=0.47\columnwidth}
\epsfig{file=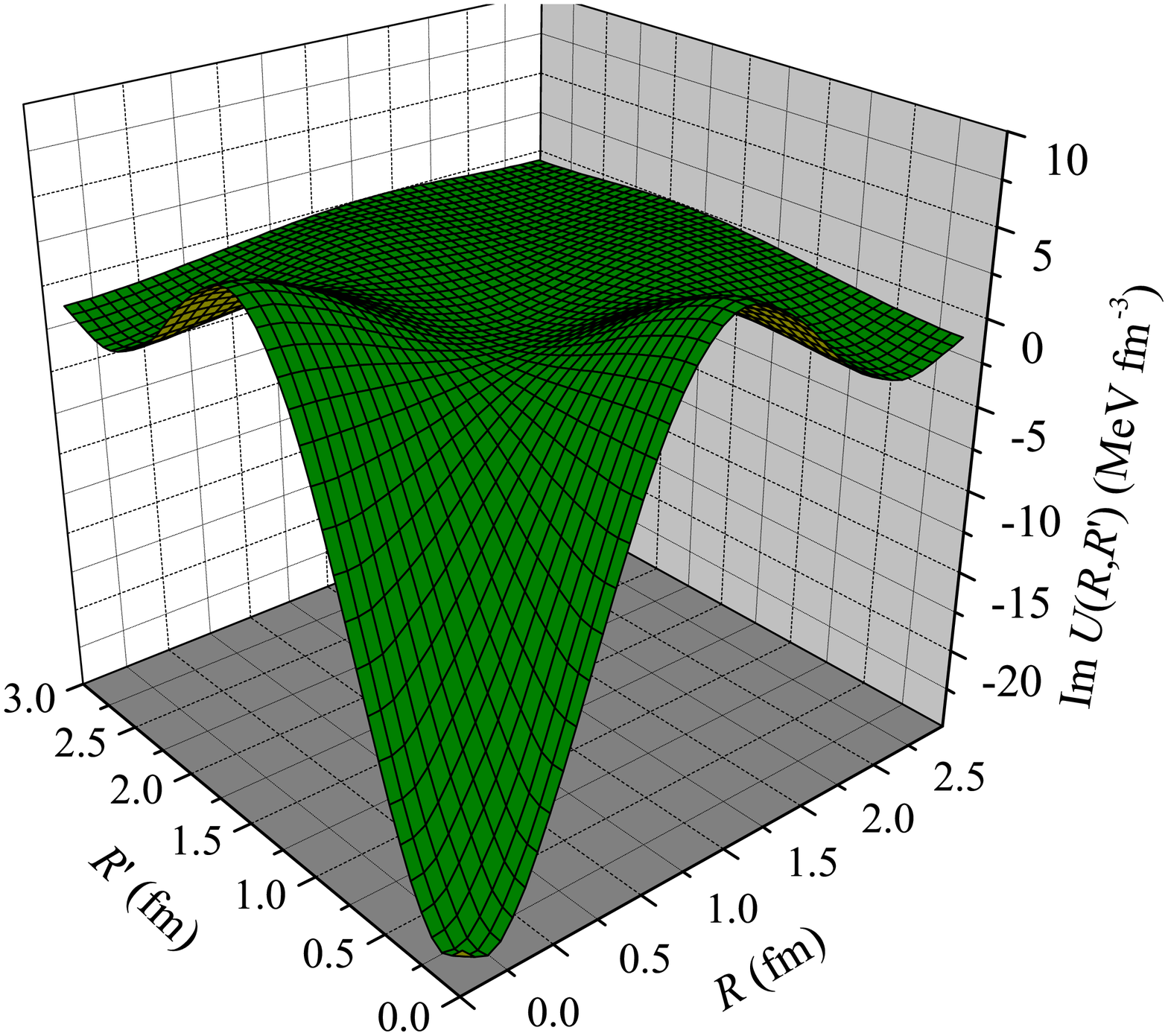,width=0.47\columnwidth}
  \caption{ \label{fig_pot} The real (left) and imaginary (right) parts of the effective
  deutron-$^{58}$Ni interaction optical potential calculated at $E_{\rm c.m.}=$80
   MeV for the total angular momentum $L=0$.}
\end{figure}

In Fig.~\ref{fig_pot} the real and imaginary parts of the
effective nonlocal optical potential for the deuteron and  the
 $^{58}$Ni nucleus interaction at incident deuteron energy $E_d=80$ MeV
calculated via the WP-approach are displayed
 for the total orbital angular momentum $L = 0$.

Besides the direct calculation of the complicated nonlocal interaction operator,
the approach described above is very convenient in those cases where
inelastic channels play a role of a
  correction  to the
elastic scattering when the main contribution comes from the folding
potential. In such a case, one can employ the ``inner'' few-body WP
basis of a rather small dimension for the calculation of the
effective potential and the ``external'' two-body WP basis of a
large dimension
 for the subsequent solution for  the resulting two-body
scattering problem with the total interaction including the folding
potential and the above effective potential \cite{K3}.

Just such a situation arises very often when the energy of collision
of a composite particle with a stable target increases. The relative
weight of the breakup channels decreases (after some characteristic
energy)
  and thus the contribution of
the effective potential taking into account the projectile
breakup  is reduced as in our above example.

\begin{figure}[h!]
 \centering \epsfig{file=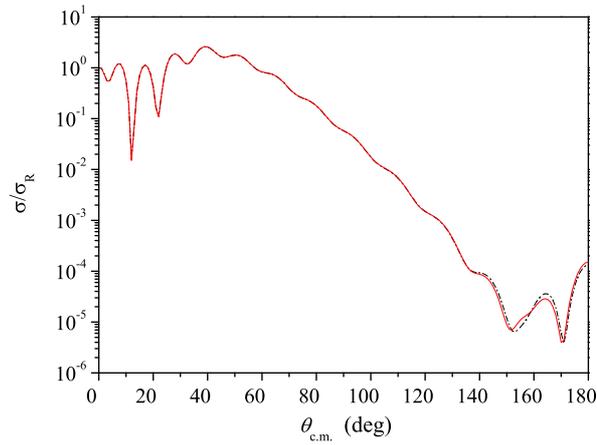,width=0.7\columnwidth}
  \caption{ \label{fig80} The differential cross section for the elastic deutron-$^{58}$Ni
  scattering at $E_{\rm c.m.}=$80 MeV calculated in the CDCC approach (dash-dotted curve)
  \cite{Moro} and  with an effective
   potential of the Feshbach type constructed in the WP approach (solid curve).}
\end{figure}

As an illustration for such an approach we present in
Fig.\ref{fig80} the differential cross section for elastic deuteron
scattering off the nucleus $^{58}$Ni at energy $E_d = $80 MeV
calculated using the above Feshbach potential.  The result of such
optical model calculation nearly coincide with the solution of a
direct three-body problem found with the CDCC approach in
ref.~\cite{Moro}. The dimensions of the Coulomb WP-basis
$|\bar{x}_i^C\rangle$ used in $Q$ projector is in five times less
than the dimension of the same Coulomb WP-basis  used in
``external'' projector $F$.

Let us summarize, with the described  technique we are able rather
easily  to calculate the theoretical non-local optical potentials
for a scattering of composite projectiles off stable targets (or
vice versa), or even for a collision  of two composite particles.

\section{Formulation and solution of the Faddeev equations in the wave-packet representation}
The most rigorous and correct formulation for the few-body scattering
problem can be attained using the Faddeev  or  Faddeev--Yakubovsky
equations. We will demonstrate below that the wave-packet approach being applied to solving such
equations gives the great advantages.

\subsection{Formulation of Faddeev equations in momentum space and
peculiar properties of their solution}
 Here we consider a solution of the Faddeev
equations   for a scattering of three identical
particles 1, 2 and 3
 with mass $m$ (nucleons). In this case the elastic scattering observables can
 be found from
a single Faddeev equation (FE) for the transition operator $\bar U$
(the so-called AGS equation),
 e.g. in the following form \cite{Gloeckle_rep}:
\begin{equation}
\bar{U}=PG_0^{-1}+PtG_0\bar{U},
\label{FEG}
\end{equation}
where  $t$ is two-body off-shell $t$-matrix in three-body space,
$G_0=(E+i0-H_0)^{-1}$ is
 the free three-body resolvent and $P$ is the
permutation operator which changes
the momentum variables from one Jacobi set to another one. For the case of three
identical particles the operator $P$ is defined as sum of
two cyclic permutations of particles:
\begin{equation}
P=P_{12}P_{23}+P_{13}P_{23}.
\end{equation}
It should be emphasized that a similar permutation operator is
included in kernels of the Faddeev equations  in the case of various
particles.

After the spin-angular expansion, the operator equation (\ref{FEG})
for each value of the total angular momentum and parity is reduced
to a system of two-dimensional integral equations in momentum space
(or to coupled two-dimensional integro-differential equations in the
configuration space). Although Faddeev \cite{faddeev} proved that
the kernels of these equations belong to the Fredholm type (i.e. the
inhomogeneous equation has a unique solution), a practical solution
of  coupled two-dimensional integral equations is a very complicated
and time-consuming task  due to complicated singular structure of
integral kernels and also the large number of coupled spin-orbital
channels.

One of problems is that kernels contain two-body off-shell
$t$-matrices $t(q,q',E)$ for different values of the total orbital
angular momentum and spin of the pair interacting particles, which
should be calculated many times at different energies.

The second problem here is that the Faddeev-type kernel at  real
energy contains singularities of two types: two-particle cuts,
corresponding to the presence of bound states in the subsystems, and
the three-body logarithmic singularity (at  the breakup threshold).
The two-body singularities are easily eliminated by the technique of
residues, while for a regularization of the three-body cut a number
of special techniques have been suggested in the previous years. As
a result,  the whole solution procedure becomes rather complicated.
Among such specific techniques the following are employed most often: \\
--  choice of a special quadrature grid points for a momenta $q'$
defined in one Jacobi set  depending on a momentum $q$ defined in
another Jacobi set
 together with the
spline interpolation of an unknown
function at iteration method of solving; \\
--  solution of the equations at the complex
 energy plane followed by an analytic continuation to the real axis; \\
--  shift for an integration  contour  from the real axis to the
  complex momentum plane.

However, the main characteristic feature of the Faddeev kernels is
the presence of the particle permutation operator $P$. The kernel of
this operator $P(p,q;p',q')$ as a function of the momenta contains
$\delta$-function and two $\theta$-functions and this results in
variable integration limits for  integrals in the Faddeev kernels
\cite{Gloeckle_rep}.
 Therefore, when replacing the integrals with
respective quadrature sums in numerical procedure  it is necessary
to apply an interpolation of the unknown solution depending on
several variables at each step of an iterative procedure. As a
result of such elaborated multi-dimensional interpolation on each
iteration step,  the interpolation procedure takes most of the
computational time of the whole computation and requires usage of
powerful supercomputers \cite{gloeckle12}).

 On contrast to this, the WP
method described here allows  to circumvent completely the above
difficulties in solving the Faddeev (and Faddeev--Yakubovsky)
equations. First of all, instead of eq.~(\ref{FEG}), one uses the
equivalent form of the Faddeev equation  for the transition operator
$U$ \cite{KPR_br,KPR_GPU}:
\begin{equation}
\label{pvg} U=Pv_1+Pv_1G_1U.
\end{equation}
In  eq.~(\ref{pvg})  $G_1$ is the three-body channel resolvent
introduced in Section 6. Due to the identity $tG_0\equiv v_1G_1$ the
kernels of the equations (\ref{FEG}) and (\ref{pvg}) are the same
and their solutions  $\bar U$ and $U$ coincide on-shell and
half-shell.

This form of the equation is especially useful in the WP
representation with basis functions corresponding to the  channel Hamiltonian $H_1$, because  in this
representation the resolvent $G_1$ has a simple analytical form
and explicitly depends only on the partitions of the continuous
spectrum of the subsystem. Thereby the need for multiple
calculations and interpolation of the off-shell $t$-matrix is
eliminated.

Further, all the singularities of the Faddeev kernel in the form
(\ref{pvg}) are concentrated in the channel resolvent $G_1$ and
they are smoothed when projecting on the WP basis (averaged by the
integration over the energy bins). Therefore the resulting matrix
equation can be solved directly for real energies.

Finally, the use of matrix of permutation in the WP basis (as
well as in any other fixed basis) completely eliminates the need
for very numerous interpolations of the required solution at each
iteration.

So, all these innovations taken together lead to great
simplification in practical solving the Faddeev-type few-body
scattering equations.

\subsection{Three-body wave-packet  basis for  $3N$ system}
To illustrate this novel technique  we consider the realistic Nd
scattering problem, where one needs to particularize the three-body
wave-packet basis for the case of three-nucleon system with tensor
$NN$ interactions. We use the following quantum numbers for the
subsystems defined in  Section \ref{three} and in the whole
three-body system  according to the $(jj)$-coupling scheme:
\begin{equation}
\alpha=\{l,s,j_{23}\};\qquad \beta=\{\lambda,I\}; \qquad
\Gamma=\{J,\pi,T\}, \label{qnum}
\end{equation}
where  $l,s$ and $j_{23}$ are the  $NN$ subsystem quantum numbers:
$l$ is an orbital momentum, $s$ is a spin and ${\bf j}_{23}={\bf
l}+{\bf s}$ is a total angular momentum of the  subsystem (the
interaction potential depends, in general, on the value of
$j_{23}$). The other quantum numbers are the following:  $\lambda$
is an orbital momentum and ${\bf
I}=\mbox{\boldmath$\lambda$}+\mbox{\boldmath$\sigma$}$ is a total
momentum of the third nucleon, where $\sigma=\half$ is its spin.
Finally, ${\bf J}={\bf j}_{23}+{\bf I}$ is a total angular momentum
of the three-body system, $T$ is a total isospin  and $\pi$ is
parity, all of them are being conserved. Let's also note that the
pair isospin $t$ can be defined by values of $l$ and $s$, because
the sum $l+s+t$ must be odd.

The two-body free WP states (\ref{pq})  should be defined for each
partial wave $l$ and $\lam$ and further they are multiplied by
appropriate spin-angular states according to (\ref{qnum}).  Thus the
free three-body basis function (\ref{xij}) can be written  in the
detailed form:
\begin{equation}
\label{xij1}
|X_{ij}^{\Ga\al\be}\rangle=|x_i^l,\bar{x}_j^\lam;\alpha,\beta:\Gamma\rangle,
\end{equation}

The three-body WP states corresponding to the channel Hamiltonian
$H_1$ are defined here as has been detailed in Section~\ref{three}.
However now one has to take into
 account all
 possible spin-angular couplings in $\{23\}$ subsystem induced by tensor
 couplings
 in $v_1$ interaction, so the two-body scattering WPs of $h_1$ sub-Hamiltonian must be defined in
the eigenchannel representation as in eq.~(\ref{mult_zi}). Then,
three-body channel WP states have the following form:
\begin{equation}
\label{si1} |Z^{\Ga\tal\beta}_{kj}\rangle
\equiv|z^{\vak}_k,{\bar x}_j^\lam;
\tal,\beta:\Gamma\rangle,
\end{equation}
where the spin-angular quantum numbers $\tal$ in $\{23\}$ subsystem
corresponds to the eigenchannel representation and differs from
those, i.e. $\alpha$, for the free WP basis. In particular, instead
of an angular momentum value $l$ this index contains
eigenchannel value $\vak$, so $\tal=\{\vak,s,j_{23}\}$ while
$\al=\{lsj_{23}\}$.

These states  are the WP analogs for the three-body  scattering
states $|\psi^{\varkappa}_{p},q;\tal,\be:\Ga\rangle$ of the channel
Hamiltonian $H_1$, where $|\psi^{\varkappa}_{p}\rangle$ is a
scattering wave function of $h_1$ sub-Hamiltonian defined in the
ER.

In Section 5, it was shown that the pseudostates obtained via the
diagonalization sub-Hamiltonian $h_1$ in a free WP basis are  good
approximations for the exact WPs $|z_{k}^{\vak}\rangle$ constructed
from the scattering wave functions for $h_1$. The  exact WPs are
related to the free WPs by a simple orthogonal
 transformation:
\begin{equation}
\label{rotation} |z_{k}^\vak,\tal\rangle=\sum_{i=1}^N\sum_{l}
O_{ki}^{\vak l}|x_{i}^{l},\al\rangle,
\end{equation}
where spin-angular parts of wave functions are taken into account as
well and the multi-index $\tal=\{\vak,s,j_{23}\}$ related to the ER
is used.

Now we have in our disposal the WP basis for the channel three-body
Hamiltonian and can apply explicit formulas (\ref{dg1}) for the
channel resolvent $G_1$ which enters eq.~(\ref{pvg}).

\subsection{Matrix analog for the Faddeev  equation in the WP basis}
In our  approach, all the operators in eq.~(\ref{pvg}) are projected
onto  3WP basis corresponding to the channel Hamiltonian $H_1$. In
other words, every operator, e.g. $U$, is replaced with its
finite-dimensional WP representation:
\begin{equation}
\label{umatr}
\mathfrak{U}^\Gamma=\sum_{\tal,\be,\tal',\be'}\sum_{k,j,k',j'}
|Z_{kj}^{\Gamma\tal\beta}\rangle \langle
Z_{kj}^{\Gamma\tal\beta}|U|Z^{\Gamma \tal'\beta'}_{k'j'}\rangle
\langle Z^{\Gamma \tal'\beta'}_{k'j'}|.
\end{equation}
As a result, one gets the matrix analog for the Faddeev equation~(\ref{pvg}) (for
 each value of $\Gamma$)
\begin{equation}
\label{m_pvg} {\mathbb U}={\mathbb P}{\mathbb V}_1+{\mathbb
P}{\mathbb V}_1 {\mathbb G}_1 {\mathbb U}.
\end{equation}
Here ${\mathbb V}_1$ and ${\mathbb G}_1$ are the matrices of the
pairwise interaction and the channel resolvent respectively, the
matrix elements of which can be found in an explicit form.

Thus, to find  the elastic scattering amplitude it is required: 1) to
calculate  matrix elements of $\mathbb P$, ${\mathbb V}_1$, ${\mathbb G}_1 $ matrices
and 2) to solve the system of algebraic equations (\ref{m_pvg}).

The matrix ${\mathbb V}_1$ of the potential  $v_1$  is diagonal in
the indices $j,j'$ of the wave-packet basis  for the free
sub-Hamiltonian $h_0^1$ and  has the block form:
\begin{equation}
\label{bv1} [{\mathbb
V}_1]^{\tal\beta,\tal'\be'}_{kj,k'j'}=\de_{\beta\beta'}\de_{jj'}\de_{ss'}\de_{j_{23}j'_{23}}
\langle z_k^\vak|v_1|z_{k'}^{\vak'}\rangle_{j_{23}}
\end{equation}
for $\tal=\{\vak,s,j_{23}\}$.
Here the subindex $j_{23}$ in the  potential $v_1$ matrix element means that it depends on
the two-body total angular momentum value. The matrix elements (\ref{bv1}) do not depend on
index $j$ and can be written with the usage of the rotation matrix
$\mathbb O$ defined in eq.~(\ref{rotation}) as:
\begin{equation}
\label{vks}
 \langle z_k^\vak|v_1|z_{k'}^{\vak'}\rangle_{j_{23}}
 =\sum_{i,i'}O_{ki}^{\vak l} O_{k'i'}^{\vak'l'} \langle
x_i^l|v_1|x_{i'}^{l'}\rangle_{j_{23}}\nonumber,
\end{equation}
where $\langle
x_i^l|v_1|x_{i'}^{l'}\rangle_{j_{23}}$ are the potential matrix elements in the free WP
basis.

The matrix of the operator $P$ in the free three-body packet basis corresponds to
the overlap between basis functions defined in different Jacobi
sets:
\begin{equation}
[{\mathbb P}^0]_{ij,i'j'}^{\al\be,\al'\be'}\equiv \langle
X_{ij}^{\Gamma\al\be}|P|X_{i'j'}^{\Gamma\al'\be'}\rangle= \langle
X_{ij}^{\Gamma\al\be}(1)|X_{i'j'}^{\Gamma\al'\be'}(2)\rangle+\langle
X_{ij}^{\Gamma\al\be}(1)|X_{i'j'}^{\Gamma\al'\be'}(3)\rangle,
\end{equation}
where the argument  1 (or 2 and 3) in the basis functions denotes a
corresponding Jacobi set. Such matrix elements can be calculated by
integration over the basis functions in momentum space:
\begin{eqnarray}
 [{\mathbb P}^0]_{ij,i'j'}^{\al\be,\al'\be'}=
\label{perm}\int_{\MD_{ij}}dpdq\int_{\MD'_{i'j'}}dp'dq'
\times\frac{P^{\Gamma}_{\al\be,\al'\be'}(p,q,p',q')}{\sqrt{d_id_{i'}\bar{d}_j\bar{d}_{j'}}},
\end{eqnarray}
where the prime at the lattice cell   $\MD'_{i'j'}$ indicates that
the cell belongs to the other (rotated) Jacobi set while the
$P^{\Gamma}_{\al\be,\al'\be'}(p,q,p'q')$ is the kernel of particle
permutation operator in a momentum space. This kernel, as is
mentioned above, is proportional to the product of a Dirac delta and
Heaviside theta functions. However, due to the integration over
momentum bins in eq.~(\ref{perm}) these singularities are averaged
over the momentum lattice cells and, as a result, the elements of
the permutation operator matrix in the WP basis are finite and
non-singular.

The detailed  technique for the calculation of the matrix element in
eq.~(\ref{perm}) is presented in ref.~\cite{KPR_GPU}.

The permutation operator matrix $\mathbb P$ in the channel 3WP basis
is expressed through the overlap matrix ${\mathbb P}^0$ for the
lattice basis function of eq.~(\ref{perm}) with  help of the
rotation matrices $\mathbb O$ if one uses  the pseudostate
approximation (\ref{rotation}) for the scattering WPs
$|z_{k}^s\rangle$:
 \begin{equation}
 \label{perm_z}
\langle
Z_{kj}^{\Gamma\tal\beta}|P|Z_{k'j'}^{\Gamma\tal'\beta'}\rangle
\approx \sum_{ii'}\sum_{l,l'}O_{ki}^{\vak l} O_{k'i'}^{\vak' l'
}[{\mathbb P}^0]_{ij,i'j'}^{\al\be,\al'\be'}.
\end{equation}

So, we have a relatively simple formulas and numerical algorithms to determine all
the quantities entering  the kernel of the matrix Faddeev equation(\ref{m_pvg}).

\subsection{Determination of observables in the WP approach}
\begin{figure}[h!]
\centering\epsfig{file=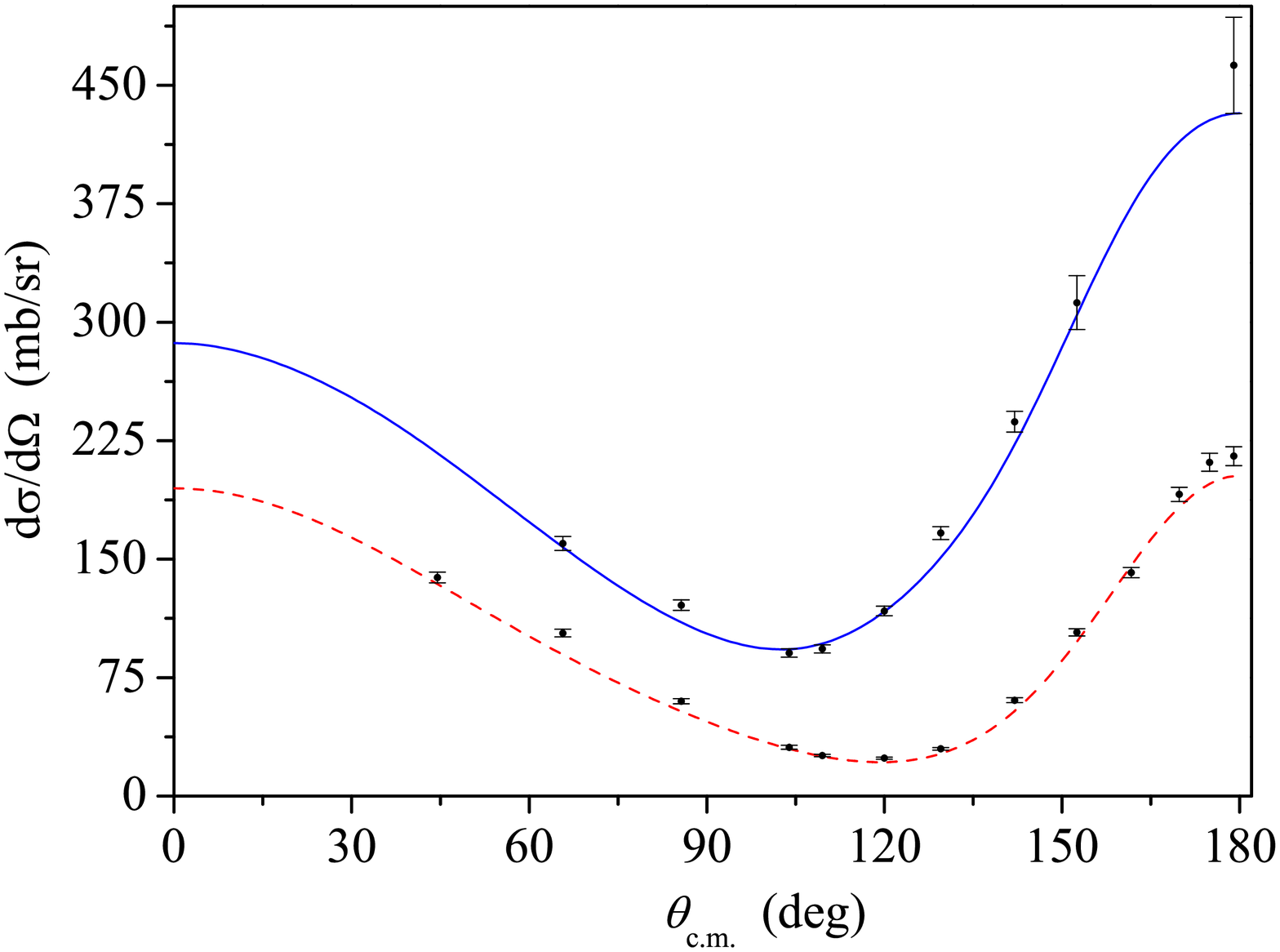,width=0.9\columnwidth} \caption{
The
 differential cross section for the elastic $nd$ scattering calculated via
the WPCD approach with the Nijmegen I NN potential for incident
neutron energies  $E_n=3$ MeV (solid curve) and $E_n=9$~MeV (dashed
curve) in comparison with the experimental data (circles)
\cite{schwartz}.\label{crosses}}
\end{figure}

The elastic on-shell  amplitude in the wave-packet representation is
calculated  as  a diagonal (on-shell) matrix element of  $\mathbb
U$-matrix \cite{KPR_br}:
\begin{equation}
A_{\rm el}^{\Gamma\al_0\be}(q_0)\approx  \frac{2m}{3q_0}
\frac{\langle
Z^{\Ga\alpha_0\beta}_{0j_0}|\mathbb{U}|Z^{\Ga\alpha_0\beta}_{0j_0}\rangle}{\bar{d}_{j_0}},
\end{equation}
 where $m$ is the nucleon mass, $q_0$ is the initial two-body momentum
and\\
$|Z^{\Ga\alpha_0\beta}_{1j_0}\rangle=|z_{1}^{\alpha_0},\bar{x}_{j_0}^\lam;\al_0,\be:\Ga\rangle$
is the 3WP basis state corresponding to the initial scattering
state. Here $|z_{1}^{\al_0}\rangle$ is the bound state of the  $NN$
pair in the initial state (the deuteron, in our case), the index
$j_0$ denotes the bin $\BMD_{j_0}$ including the on-shell momentum
$q_0$ and $\bar{d}_{j_0}$ is a momentum width of this bin.

In Fig.~\ref{crosses}  the  differential cross sections for $nd$
elastic scattering found with the Nijmegen I $NN$ potential
\cite{nijm} in the WP approach are represented for different
incident neutron energies in comparison with experimental data
\cite{schwartz} (circles). It is evident from the Figure that
agreement with the data is very well.

\begin{figure}[h!]
\centering \epsfig{file=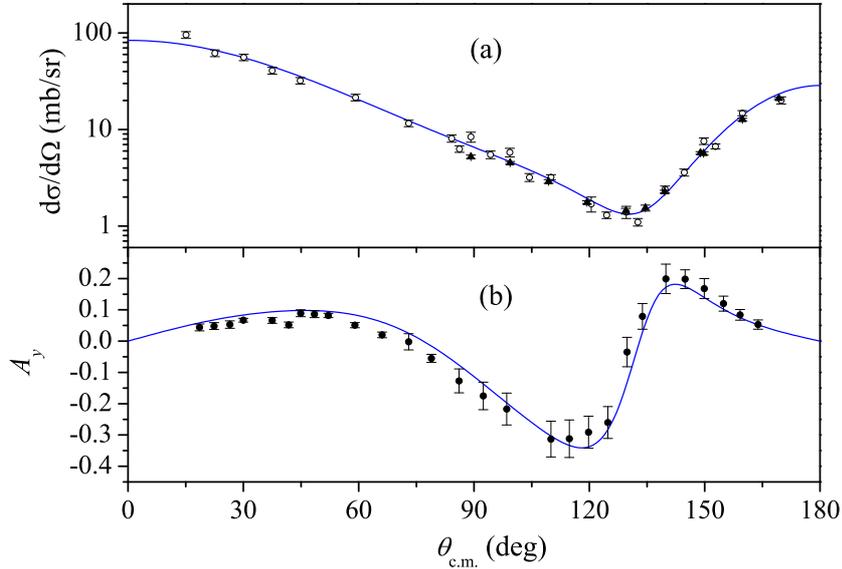,width=0.9\columnwidth} \caption{
The elastic $nd$ scattering differential cross section (a) and the
neutron vector analyzing power $A_y$ (b)  at 35~MeV obtained within
the WP approach (solid line). The experimental sets are the
following: $pd$ data at 35 MeV  \cite{pd_exp} (full circles),  $nd$
data at 36 MeV \cite{nd36} (empty circles) and $nd$ data 35 MeV
\cite{nd35} (triangles) correspondingly .\label{ay} }
\end{figure}

 In  Fig.~\ref{ay} the  comparison is given
for the neutron vector analyzing powers $A_y$ for the elastic $nd$
scattering at 35~MeV.  Here the WP basis with dimension $N\times
\bar{N}=100\times 100$ has been used and the partial waves with
the total angular momentum up to $J\le 17/2$ have been taken into
account.
\subsection{Treatment of the three-body breakup in the WP-approach}
Now let's pass to a treatment for the three-body breakup using the
WP approach. One can show \cite{KPR_br} that the breakup amplitude
may be defined as a matrix element of {\em the same transition
operator} $U$ satisfying eq. (\ref{pvg}), not only as its diagonal
element on initial states, but also for the transition to the
continuum state of the channel Hamiltonian $H_1$:
\begin{equation}
\label{t_sig} T(p,q)\sim \frac{\langle
z_1^{\al_0},q_0;\al_0,\be:\Ga
|U|\psi_p^{\al(+)},q;\al,\be:\Ga\rangle}{pqq_0},
\end{equation}
where $|\psi_p^{\al(+)}\rangle$ is the scattering function for the
Hamiltonian $h_1$ corresponding to the outgoing boundary condition,
$p,q$ are the final momenta of the subsystem $\{23\}$ and the third
nucleon correspondingly,  $q_0$ is the initial momentum of the third
nucleon which are interrelated to each other by the energy
conservation $\ep_1^*+\frac{3q_0^2}{4m}=\frac{p^2}{m}+
\frac{3q^2}{4m}$.

Thus, in the WP-approach,  the breakup amplitudes can be defined
quite similarly to a matrix element for the elastic scattering
transition operator $U$ with evident replacement of the  $NN$
bound-state wavefunction with the exact scattering functions for the
$NN$ sub-Hamiltonian \cite{KPR_br} (or the corresponding WP state).

As an illustration, consider the case of semi-realistic $s$-wave
$NN$ interactions MT~III. In this calculation one takes $l=\lam=0$
and one has only $NN$ spin quantum number $s$ to distinguish
different spin-angular channels: $\al_0=s=1$ for the initial
channel, $\al=s=0,1$ for the final channel and the three body
quantum number $\Ga$ is defined by the total spin value $\Sigma$. We
show here the results for the hyperspherical breakup amplitude which
defines an asymptotic behavior of a  Faddeev component of the wave
function and is related to the breakup amplitude (\ref{t_sig}) as
follows:
\begin{equation}
\label{acal_t} {\cal A}^{\Sigma}(\theta)=\frac{4\pi m}{3\sqrt{3}}q_0
K^4 e^{i\pi/4}T^{\Sigma s}(p,q),\ \theta=\arctan(\frac{\sqrt3q}{2p})
\end{equation}
where  $\theta$ is the hyperangle in momentum space. In the WP
approach the breakup amplitude $T$ is defined by  non-diagonal
elements of the transition operator $\mathfrak U$ (see details in
ref.~\cite{KPR_br}):
\begin{eqnarray}
T^{\Sigma s}(p,q)\approx e^{i\delta(p^*_k)}\frac{{\mathbb T}^{\Sigma
s}_{1j_0,kj}}{p_k^*q_j^*q_0},\nonumber\\
{\mathbb T}^{\Sigma s}_{1j_0,kj}\equiv\frac{\langle
Z^{\Sigma1}_{1j_0}|\mathfrak{U}^{\Sigma}|Z^{\Sigma
s}_{kj}\rangle}{\sqrt{\bar{d}_{j_0}d_k^s\bar{d}_j}} ,\quad
\begin{array}{c}
q_0\in \BMD_{j_0},\\
q\in \BMD_j,\\
p\in \Delta_k^s,\\
\end{array}
\end{eqnarray}
where $|Z^{\Sigma1}_{1j_0}\rangle$ is the WP basis state
corresponding to the initial state: index 1 denotes the bound state
of the $NN$ pair (deuteron) and index $j_0$  denotes the
``on-shell'' $q$-bin $\BMD_{j_0}$, while $|Z^{\Sigma s}_{kj}\rangle$
defines the  state of the channel Hamltonian $H_1$ ``excitation'' to
which corresponds to a breakup process. Here $\delta(p^*_k)$ is the
$s$-wave phase shift corresponding to the $NN$ pair interaction at
 energy $\ep_k^{*s}$, $p_k^*=\sqrt{m\ep_k^{*s}}$ and
$q^*_j=\half[q_{j-1}+q_j]$ are momenta corresponding to $\Delta_k^s$
and $\BMD_j$ bins respectively and the $d_k^s$ is the momentum width
of the $\Delta_k^s$ bin.

\begin{figure}[h!]
\centering \epsfig{file=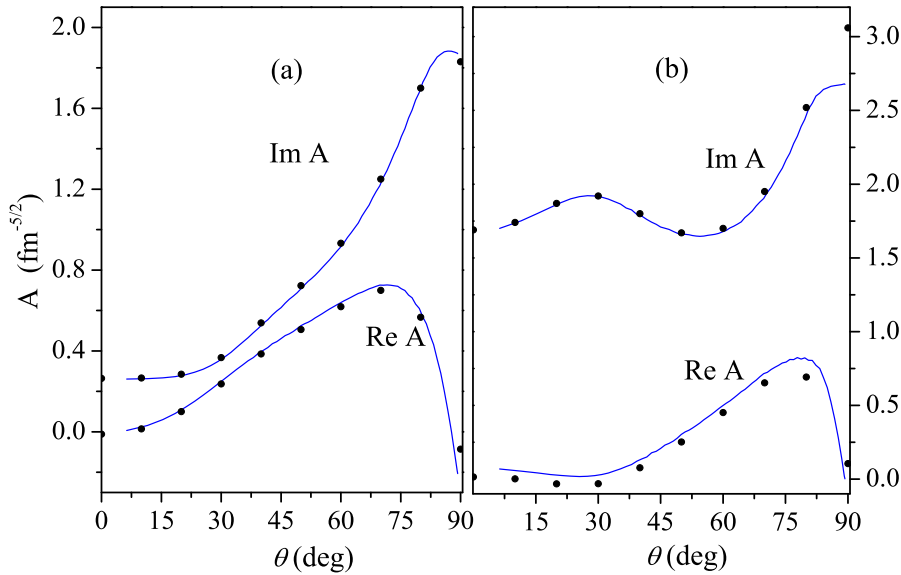,width=0.9\columnwidth}
\caption{\label{fig_br}A comparison of the hyperspherical
 breakup amplitudes ${\cal A}$
  for the spin-quartet (a) and spin-doublet
   (b) channels obtained within WP technique (solid curves)
   and in the conventional Faddeev calculations \cite{friar} (full circles).}
\end{figure}

A comparison for the hyperspherical
breakup amplitudes defined in the WP approach with results of the
benchmark solution of the Faddeev equation \cite{friar} is displayed  in Fig.~{\ref{fig_br}}.

\subsection{Some features of our algorithm for solution of
Faddeev equations in the WP approach. Realization through the GPU
parallel computations} Using the WPCD approach, we have reduced the
practical solution of the Faddeev equation to solving the system of
algebraic equations (\ref{m_pvg}). Thus, the WPCD algorithm for
solving system  consists of the following main steps:
\begin{enumerate}
\item Construction of free WP bases $\{|x_{i}\rangle,|\bar{x}_{j}\rangle\}$,
 calculation of the potential matrix
$\langle x_i^\alpha|v_1|x_{i'}^{\alpha'}\rangle $, diagonalization
of  two-particle sub-Hamiltonian matrices for each value of $l$ and
$s$
 and  finding the energies of pseudostates and their functions in the WP
 representation, i.e. matrices $ O_{ki}^{\vak l}$.
\item Calculation of the permutation matrix ${\mathbb P}^0$ in the lattice
     WP basis (\ref{perm}).
\item Calculation of the channel resolvent matrix $ {\mathbb G}_1(E)$
in the WP basis corresponding to the channel Hamiltonian $ H_1$ (the diagonal matrix).
\item Solution of the system of algebraic equations (\ref{m_pvg}) and
the determination of the elastic and breakup amplitudes.
\end{enumerate}

Point 1  provides two-particle input for a solution of the
three-body problem. In our WP approach, the input is obtained by a
single (for each value of $\alpha$) diagonalization of the
two-particle Hamiltonian.  The result of these diagonalizations can
be used to solve the scattering problem at different energies $E$.
Whereas in the standard approach to solve the Faddeev equations one
has to calculate  two-particle off-shell $t$-matrices many times for
each value of $E$.

Point 2 is the key to simplify the whole solution of the
Faddeev-type equations. The use of a finite-dimensional
approximation for the permutation operator, i.e. replacing it with a
fixed matrix (in any basis), avoids  the need for  numerous and
time-consuming multi-dimensional interpolations of a current
solution during the iteration process. In the standard approaches,
these interpolations take the most of the computing time. Although
in the calculation of matrix elements of permutation operator in our
approach also meets some difficulties --- multiple integrals with
variable limits, however these matrix elements are computed with
simple functions.
 Nevertheless, the calculation of the permutation matrix ${\mathbb P}^0$ takes
the major part of computing time in
 our algorithm in sequential execution on the CPU. But it should be stressed that the matrix ${\mathbb P}^0$
is independent of energy and therefore being calculated at once, it
can be used to solve scattering problem at so many energies as we
wish. Note also that due to energy conservation
($p_1^2/m+3q_1^2/(4m) = p_2^2/m+3q_2^2/(4m)$, where subscripts 1 and
2 indicate the different Jacobi sets, the permutations matrix
${\mathbb P}^0$ is very sparse -- only about 1~\% its matrix
elements are nonzero.

Point 3 is based on the main feature of the WP representation. It is
just the advantage of the WP basis  that  the resolvent matrix
${\mathbb G}_1(E)$ for the channel Hamiltonian $H_1$  is  diagonal,
does not depend explicitly on interaction potentials and its
elements are determined only by the partitions of two-dimensional
bins and the total energy $E$ using explicit formulas.

The only difficulty  in the  practical  solution of the matrix
equation (\ref{m_pvg}) is its high dimensionality --- that is the
price for the elimination of the basic difficulties of a standard
approach to solving the Faddeev equation. Solution of an algebraic
system of such high dimensionality is an untrivial problem even for
a supercomputer. But in our approach we do not need in solution for
the whole system.
 Indeed, to find the
elastic and breakup amplitudes one needs only on-shell matrix
elements of the transition operator. Each of these elements can be
found  by means of a simple iteration procedure (without complete
solving the matrix equation (\ref{m_pvg})) with subsequent summation
of the iterations via the Pade-approximant technique. Note that the same
Pade-technique is used in the standard approach to the solution of the
Faddeev equations \cite{Gloeckle_rep}.

\begin{figure}[h!]
 \centerline{\epsfig{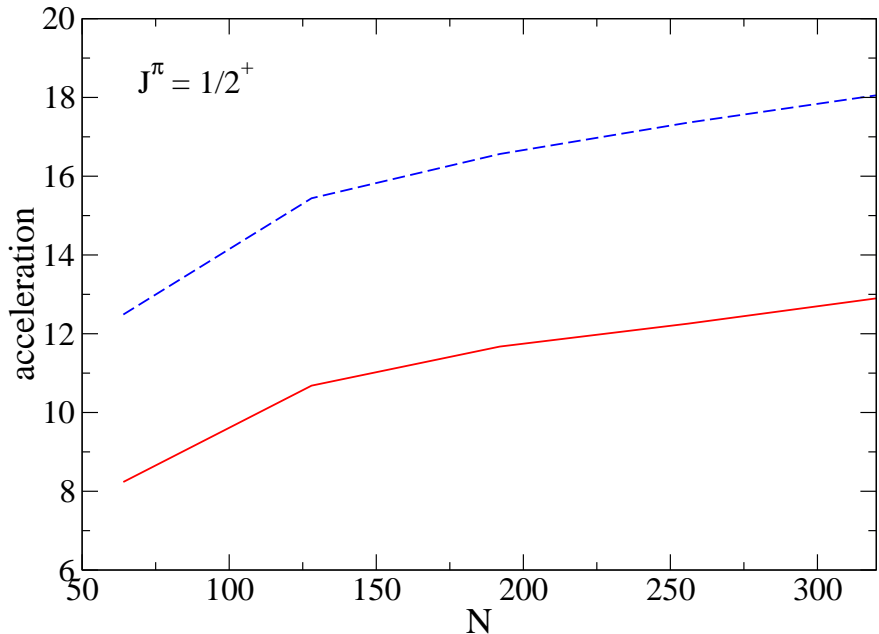}}
\caption{  The dependence of the GPU-acceleration on dimension of
the basis $N\times \bar{N}$, here $\bar{N}=N$, for  the realistic
$nd$ scattering problem with Nijmegen I $NN$ potential at
$J=\frac12^+$: dashed line shows the acceleration for the step 2
(calculation of the permutation matrix ${\mathbb P}^0$),
 solid line -- the  acceleration for the complete solution (see the main text).
 \label{accel_GPU}}
\end{figure}

As was  mentioned earlier, the main computational effort in our
case is spent on the calculation of the free WPs overlap matrix
$\mathbb P^0$.
 Because these elements are computed independently from each other,
the algorithm is very appropriate for parallelization and
implementation on multiprocessor systems, in particular
 on a graphics processing unit (GPU). We have adopted
the multi-thread GPU algorithm for a fully realistic calculations
for $nd$ scattering above the breakup threshold and have attained
the great (more than tenfold) acceleration  in a comparison with the
  calculation on the same PC without using GPU~\cite{KPR_GPU}.
The Fig.~\ref{accel_GPU} illustrates dependence of the
GPU-acceleration in the solution of the Faddeev equation with a
realistic $NN$ interaction on the dimension of the WP basis. The
distinctive feature of the GPU-calculations is that the acceleration
grows  with increasing the basis dimension which is clearly seen in
Figure.

\subsection{Construction of an effective potential for $nd$ scattering}
It has been suggested  \cite{Gloeckle_rep}, that the  $nd$
scattering problem can be solved by constructing an  effective  $nd$
interaction potential.

Indeed, let us consider again the equation for the transition
operator (\ref{pvg}) and divide the channel resolvent $G_1$ into two
parts (according to the projectors $F$ and $Q$ introduced in the
Section 7)
\begin{equation}
G_1=G_{1F}+G_{1Q}.
\end{equation}
It is evident that this division corresponds just to bound-continuum
and continuum-continuum parts of the channel resolvent, i.e.
$G_{1F}=G_1^{\rm BC}$ and $G_{1Q}=G_1^{\rm CC}$, where the first
operator contains channel states including deuteron bound state
$|z_1\rangle$ .

Then, instead of a single equation (\ref{pvg}), one can derive two
equations:
\begin{eqnarray}
U=Pv_1+Pv_1G_1^{\rm BC}{\cal V}\label{cal_u}\\
{\cal V}=Pv_1+Pv_1G_1^{\rm CC}{\cal V}\label{calv}.
\end{eqnarray}
It is easy to see that the first equation essentially determines
the elastic amplitude and is an  equation for two-particle
scattering, while the second equation defines just the effective
interaction potential of deuteron as  whole with an incident
neutron and takes into account inelastic processes.

In the traditional approach, the solution of equation (\ref{calv})
is practically very hard, because it requires knowledge of the
channel resolvent in the subspace orthogonal to the bound state
(i.e. its
 continuum-continuum part).  In the WP approach, the solution of
this equation  presents no difficulty, since the channel resolvent
$G_1^{\rm CC}$ is well known here. Then the method for solving the
problem of $nd$ scattering by means of effective potentials can be
represented schematically as follows:
\begin{itemize}
\item[(i)] introduction of the 3WP basis for the channel Hamiltonian and construction
of an analytical finite-dimensional representation for $G_1^{\rm
CC}$ operator; \item[(ii)] solution of eq.~(\ref{calv}) in the above
WP representation; \item[(iii)] solution of eq.~(\ref{cal_u}) in a
two-body free WP basis using a matrix form of $\cal V$ operator
obtained at the previous stage.
 \end{itemize}

 \section{Summary}
We described in the present paper a general technique for the continuum discretization
in few-body scattering problem based on projection of scattering operators  and respective
wave functions onto the discrete basis of stationary wave packets.
 The basic idea behind the approach is
 a similarity (within some restricted space) of exact
non-normalized scattering states of the Hamiltonian and the
respective discrete and normalized wave-packet basis states. So
that, such a WP projection allows us to transform the complicated
singular multi-dimensional integral equations describing scattering
(like general Lippmann--Schwinger or Faddeev--Yakubovsky equations)
to regular matrix equations which can be solved directly within
computational procedures similar to those used in bound-state
calculations.

This novel approach has a few characteristic features which allow to
simplify drastically the whole solution of  few- and many-body
scattering equations.

First, owing to some averaging the integral kernels over the
momentum cells, all their complicated moving singularities  are
smoothed out and as a result one gets the simple matrix equations
with finite matrix elements. This allows one to solve the resulting
matrix equations directly on real energy axis without any contour
deformation, continuation to complex energy plane or special
interpolation procedures.

Second, instead of fully off-shell $t$-matrix at many energies
entering the integral kernel in a conventional approach, one uses
the initial potential (the matrix of which is calculated easily) and
the matrix of the channel resolvent  which is found
 by means of explicit formulas.

Third, the exact scattering wave packets in pair subsystems are
treated as usual normalized excited states.  This makes  possible to
consider a  three-body breakup process as an inelastic scattering
   to normalized pseudostates. Such a replacement simplifies greatly
the description of  three- and few-body breakup processes.

Fourth, due to a specific matrix structure of resulting matrix
equations one can organize the massively-parallel computing with use
of   the graphics processing unit. This makes it possible to carry
out all the computations via many thousands of threads on the GPU
inside a desktop PC and leads to real ultra-fast calculations for
the scattering problems in few-body systems.

The developed wave-packet approach to solving few-body scattering
problems is universal and may be used in different branches of
nuclear, atomic and chemical physics, in quantum statistics and
nuclear matter theory. Also it can be directly generalized to a
relativistic case.

{\bf Acknowledgments} The authors thank Profs. A.K.~Motovilov and
V.V.~Pupyshev
 for the discussions
 of  mathematical aspects of our approach.  This work has been supported
 partially by the Russian Foundation for Basic Research,
 grant No.13-02-00399.

\appendix
\section{Explicit formulas for the channel resolvent eigenvalues}

\subsection{Eigenvalues of two-body resolvent}
Two-body free resolvent eigenvalues  for the case of momentum WPs
($f(q)=1$) are defined by the general formula (\ref{ri}) and thus they
have the following form:
\begin{eqnarray}
  g_i(E)=
 \frac{\mu}{qd_i} \Big\{ \ln\left|\frac{q-q_{i-1}}{q-q_i}\right| +
\ln\left|\frac{q+q_{i}}{q+q_{i-1}}\right|     - \nonumber \\
i\pi\left[\theta(q-q_{k-1})-\theta(q-q_{k})\right] \Big\},
\end{eqnarray}
where $q=\sqrt{2\mu E}$ and the combination of the Heaviside
$\theta$-functions means that the imaginary parts of the eigenvalues
don't vanish only in a single  interval to which the respective
on-shell momentum value  belongs: $q\in\MD_k$.

The energy  averaged eigenvalues are defined by integral
\begin{equation}
g_i^{k}\equiv
\frac{1}{D_k}\int_{\MD_k}g_i(E)\frac{q}{\mu}{\rmd}q,\quad E\in
\MD_k.
\end{equation}
Explicit formulas for the $q$-packet case now are the following:
\begin{equation}
\label{evg_av} g_i^{k}=\frac{1}{D_{k}d_{i}}\left[
Q_{ki}^{(+)}-Q_{ki}^{(-)}\right]-\frac{i\pi}{D_{k}}\de_{ik},
\end{equation}
where
\[
 Q_{ki}^{(\pm)}=\!
 \sum_{k'=k-1}^k\sum_{i'=i-1}^i (-1)^{k-k'+i-i'}[q_{k'}\!\pm q_{i'}]
 \ln|q_{k'}\!\pm q_{i'}|.
 \]

 In the case of the energy FPs with weight function
 $f(q)=\sqrt{\frac{q}{\mu}}$ the  formula for the free resolvent eigenvalue is the following:
 \begin{equation}
 \label{pg0}
 g_i(E)=\frac{1}{D_i}\ln\left|\frac{E-\ce_{i-1}}{E-\ce_i}\right|-
 \frac{i\pi}{D_k}\left[\theta(E-\ce_{k-1})-\theta(E-\ce_{k})\right],
 \end{equation}
 while for the averaged one
\begin{equation}
g_i^k=\frac{1}{D_kD_i}W_{ki}-\frac{i\pi}{D_k}\de_{ik},
\end{equation}
where
\[W_{ki}=\sum_{k'=k-1}^k\sum_{i'=i-1}^i (-1)^{k-k'+i-i'}[E_{k'}-E_{i'}]
 \ln|E_{k'}-E_{i'}|.
 \]

\subsection{Eigenvalues of three-body channel resolvent}
The eigenvalues for the case of energy WPs defined in (\ref{dg1}a)
for the BC part of indices  are the following
\begin{equation}
\left.
\begin{array}{l}
\displaystyle
 {\rm Re}[G^{\Gamma\tal\be}(E)]_{kj}=
\frac{1}{\bar{D}_j}\ln\left|\frac{\ce_{j-1}+\ep_{k}^*-E}
{\ce_{j}+\ep_{k}^*-E}\right|,
\\ \displaystyle
 {\rm Im}
[G^{\Gamma\tal\be}(E)]_{kj}= - \frac{ \pi}{\bar{D}_j}\left\{
\theta(\ce_j+\ep_k^*-E)-
\theta(\ce_{j-1}+\ep_k^*-E)\right\}\\
\end{array} \right\}.
\label{bc}\end{equation} These BC eigenvalues have the same
functional form as eigenvalues of the two-body free resolvent given
in (\ref{pg0}). The only difference is that bound points $\ce_j$ of
partition of the free sub-Hamiltonian $h_0^1$ continuum are shifted
to eigenvalues  $\ep_k^*$ of the sub-Hamiltonian $h_1$.

The real parts of the CC-parts of the channel resolvent have the
following form :
\begin{eqnarray}
{\rm Re}[G^{\Gamma\tal\be}(E)]_{kj}=
\frac{1}{D_{k}^{\vak}\bar{D}_j}\Big\{ (\De +\De_{-}) \ln\left|\De
+\De_{-}\right| +
 (\De -\De_{-})\ln\left|\De -\De_{-}\right|\Big\}-\nonumber\\
-\frac{1}{D_{k}^{\vak}\bar{D}_j}\Big\{ (\De +\De_{+}) \ln\left|\De
+\De_{+}\right| +
 (\De -\De_{+})\ln\left|\De -\De_{+}\right|
\Big\}, \label{ccr}
\end{eqnarray}
 where:
\[
\De\equiv \epsilon_k^{*\vak}+\ce_j^{*}-E,\quad
\De_{-}\equiv\frac{D_k^{\vak}-\bar{D}_j}2,\quad
\De_{+}\equiv\frac{D_k^{\vak}+\bar{D}_j}2
\]
and $D_k^{\vak}$, $\bar{D}_j$ are energy widths of bins for
sub-Hamiltonians $h_1$ and $h_0^1$ respectively.

 The imaginary parts of
 the CC-eigenvalues also have an explicit analytical form
 \begin{eqnarray}
 {\rm Im}[G^{\Gamma\tal\be}(E)]_{kj}=
-\frac{\pi}{D_{k}^{\vak}\bar{D}_j} \Big\{
(\De+\De_{+})\theta(\De+\De_{+})+(\De-\De_{+})\theta(\De-\De_{+})-\nonumber\\
(\De+\De_{-})\theta(\De+\De_{-})-(\De-\De_{-})\theta(\De-\De_{-})
\Big\}. \label{cci}
\end{eqnarray}

\end{document}